\documentclass{article}

\usepackage{arxiv}
\usepackage{amsmath,amssymb}
\usepackage{amsfonts}
\usepackage{bm}
\usepackage{mathtools}

\usepackage{graphicx}
\usepackage{booktabs}
\usepackage{hyperref}
\usepackage{tabularx}
\usepackage{authblk}
\usepackage{float}
\usepackage{subcaption}
\usepackage{tikz}
\usetikzlibrary{positioning,arrows.meta,shapes.geometric,fit}
\usepackage{array}

\title{Projection-Domain Sensitivity Analysis of Vertebral DRRs Under Intrinsic Calibration Perturbation}

\author[1]{Lin Li \thanks{Corresponding author: lli15@tulane.edu}}
\author[2]{Chaochao Zhou}
\author[3]{Benjamin Aubert}
\author[4]{Junlin Guo}
\author[5]{Junchao Zhu}

\affil[1]{Department of Mathematics, Tulane University}
\affil[2]{Department of Computer Science, University of Illinois Urbana-Champaign}
\affil[3]{EOS, EOS Imaging Inc.}
\affil[4]{Department of Electrical and Computer Engineering, Vanderbilt University}
\affil[5]{Department of Computer Science, Vanderbilt University}

\begin{document}

\maketitle

\begin{abstract}

Accurate geometric calibration is a fundamental requirement for fluoroscopy-guided spinal imaging, digitally reconstructed radiograph (DRR) generation, and 2D--3D vertebral registration. While calibration quality is traditionally evaluated using reconstruction-domain metrics such as reprojection error and reconstruction accuracy, the impact of calibration uncertainty on projection-domain consistency remains insufficiently understood. In this work, projection-domain consistency refers to the stability of DRR projection appearance, projected anatomical geometry, and downstream 2D--3D registration performance under changes in calibration parameters.

This paper presents a synthetic framework for analyzing the sensitivity of vertebral fluoroscopic projections to intrinsic calibration perturbations. Using CT-derived vertebral models and controlled cone-beam imaging geometry, DRRs are generated under both ground-truth and perturbed intrinsic calibration parameters while maintaining fixed anatomy and acquisition pose. Projection-domain effects are quantified using anatomical landmark displacement, contour-based distance measures, silhouette overlap metrics, image similarity analysis, and landmark-based 2D--3D registration accuracy in both anterior--posterior (AP) and lateral (LAT) views.

Experimental results demonstrate that relatively small intrinsic calibration perturbations can produce measurable changes in vertebral projection geometry, contour morphology, landmark localization, and DRR appearance despite unchanged anatomy and imaging pose. Sensitivity is strongly view dependent, with LAT projections exhibiting substantially greater contour deformation, silhouette degradation, and anatomical displacement than AP projections. Furthermore, synthetic 2D--3D registration experiments show that projection inconsistencies induced by intrinsic calibration perturbations propagate to downstream registration error, particularly in rotational alignment accuracy.

These findings highlight the limitations of relying solely on reconstruction-based calibration evaluation and demonstrate that projection-domain consistency provides complementary information relevant to DRR generation and fluoroscopy-guided registration. The proposed framework offers a controlled methodology for characterizing geometric sensitivity and calibration robustness in vertebral imaging systems.

\end{abstract}


\section{Introduction}

Fluoroscopy-guided 2D--3D registration plays an increasingly important role in spine surgery, robotic navigation, image-guided intervention, and intraoperative localization. By aligning preoperative CT data with intraoperative fluoroscopic images, these systems enable vertebral localization, surgical guidance, implant placement, and navigation with improved accuracy. Central to the success of these workflows is accurate imaging geometry calibration, which directly affects DRR generation, projection fidelity, and registration performance.

Most calibration methodologies evaluate geometric accuracy using reconstruction-domain measures such as reprojection error, triangulation accuracy, or volumetric reconstruction quality. Low reconstruction error is often interpreted as evidence that the estimated imaging geometry is sufficiently accurate for downstream applications. However, many fluoroscopy-guided registration algorithms operate primarily in the projection domain, where registration quality depends on the consistency between acquired radiographs and synthesized DRRs rather than reconstruction accuracy alone.

Recent studies \cite{2D-3DregistrationsurveyLin, li2026intrinsictolerancecarmimaging} have suggested that fluoroscopic calibration uncertainty can affect the underlying imaging geometry used for DRR generation and 2D--3D registration. In this study, calibration variation specifically refers to controlled perturbations of intrinsic calibration parameters, including effective focal scale and piercing point location, while the extrinsic acquisition pose is held fixed. This motivates the central question of this work: how sensitive are vertebral DRR appearance, projected anatomical geometry, and downstream 2D--3D registration accuracy to intrinsic calibration variation?

This question is particularly relevant for spinal imaging. Vertebral anatomy contains complex three-dimensional structures, overlapping bony features, and limited fluoroscopic viewpoints. Small changes in imaging geometry can alter vertebral contours, pedicle appearance, endplate alignment, and local projection morphology. These differences may influence image similarity metrics, registration convergence, and ultimately localization accuracy. Because DRR-based registration depends fundamentally on projection-domain consistency, understanding the relationship between calibration perturbation and projection appearance is essential.

Despite extensive research in fluoroscopic calibration, DRR generation, and medical image registration, comparatively little attention has been devoted to quantifying how calibration perturbations influence vertebral projection consistency. Existing studies primarily focus on reconstruction accuracy, rendering quality, similarity metrics, or registration optimization, while the direct relationship between calibration variation, projection deformation, and registration behavior remains insufficiently characterized.

In this work, we present a synthetic framework for analyzing projection-domain sensitivity to intrinsic calibration perturbations using CT-derived vertebral models and digitally reconstructed radiographs. Controlled perturbations are introduced into the intrinsic calibration matrix while maintaining fixed anatomy and imaging pose, enabling isolation of projection changes caused solely by calibration variation. Projection consistency is evaluated using landmark displacement, contour deformation, silhouette overlap, and DRR similarity measurements in both anterior--posterior (AP) and lateral (LAT) imaging geometries. In addition, synthetic 2D--3D registration experiments are performed to investigate how calibration-induced projection inconsistencies affect downstream rigid registration accuracy.

The overarching objective of this study was to investigate whether intrinsic calibration perturbations can substantially affect projection-domain consistency and downstream 2D--3D registration accuracy, even when the underlying anatomy and acquisition geometry remain unchanged. We hypothesized that small perturbations in intrinsic calibration parameters would produce view-dependent changes in vertebral projection appearance, with lateral projections being more sensitive than AP projections because of overlapping anatomy and compressed depth geometry. We further hypothesized that these projection-domain inconsistencies could degrade registration performance, particularly for rotational alignment estimates.

The primary contributions of this work are summarized as follows:

\begin{enumerate}
\item We introduce a controlled synthetic framework for analyzing projection-domain sensitivity to intrinsic calibration perturbations in vertebral fluoroscopic imaging.

\item We systematically quantify projection consistency using complementary landmark, contour, silhouette, and image-similarity metrics.

\item We characterize view-dependent sensitivity and demonstrate that lateral vertebral projections are substantially more sensitive to intrinsic calibration variation than anterior--posterior projections.

\item We establish a quantitative link between intrinsic calibration perturbation, projection-domain inconsistency, and downstream 2D--3D registration error.

\item We demonstrate that projection-domain analysis provides complementary information beyond conventional reconstruction-based calibration evaluation for fluoroscopy-guided vertebral imaging systems.

\end{enumerate}

\section{Related Work}

\subsection{Fluoroscopy-Guided 2D--3D Registration}

2D--3D registration aims to align preoperative volumetric anatomy with intraoperative radiographic images through optimization of geometric transformation parameters. Existing registration approaches include intensity-based, feature-based, contour-based, and gradient-based formulations \cite{maintz1998survey,zitova2003image,markelj2012review, wang2026visual}. In fluoroscopy-guided interventions, digitally reconstructed radiographs (DRRs) synthesized from CT are commonly matched to acquired X-ray images using similarity metrics such as normalized cross-correlation (NCC), mutual information (MI), and gradient correlation \cite{penney1998comparison,viola1997alignment}.

In spine surgery applications, vertebral 2D--3D registration has been investigated for vertebral localization, surgical navigation, robotic guidance, and prevention of wrong-level surgery \cite{otake2012automatic, suh2025pedicle}. However, vertebral registration remains challenging because of overlapping bony anatomy, limited fluoroscopic viewpoints, low soft-tissue contrast, and strong sensitivity to imaging geometry. Most prior studies focus primarily on optimization strategies, feature extraction, or computational efficiency while assuming accurate fluoroscopic calibration.

Because DRR-based registration fundamentally depends on projection-domain consistency, inaccuracies in imaging geometry may directly influence anatomical appearance, similarity metrics, and optimization convergence. Nevertheless, the relationship between calibration ambiguity and vertebral projection consistency remains insufficiently explored.

\subsection{Geometric Calibration in Fluoroscopic Imaging}

Accurate geometric calibration is fundamental for fluoroscopy-guided reconstruction, navigation, and image-guided intervention. Conventional C-arm calibration methods estimate intrinsic and extrinsic imaging parameters using fiducial phantoms, calibration objects, geometric constraints, or self-calibration strategies \cite{navab1998camera,siewerdsen2008cone}. Calibration quality is typically evaluated using reprojection error or reconstruction-domain accuracy metrics.

Previous work has demonstrated that calibration uncertainty can influence geometric fidelity and navigation precision in cone-beam CT and fluoroscopy-guided procedures \cite{siewerdsen2008cone,zbijewski2011volume, vanessaKAN-DDPM}. However, most existing calibration frameworks implicitly assume that accurate reconstruction guarantees sufficient projection consistency for downstream DRR-based registration tasks.

In practical imaging systems, calibration ambiguity may arise because multiple parameter combinations produce similar reprojection or reconstruction accuracy. Although such solutions may appear equivalent in the reconstruction domain, they can generate substantially different image projections, especially in anatomically complex structures such as vertebrae. Understanding the projection-domain consequences of calibration ambiguity is therefore essential for reliable fluoroscopy-guided registration.

\subsection{DRR Generation and Projection Similarity}

DRRs are widely used in image-guided surgery and fluoroscopy-guided registration to simulate X-ray projections from volumetric CT data. Classical DRR generation approaches include ray-driven projection, attenuation-based rendering, and Siddon-style ray tracing methods \cite{siddon1985fast}. Accurate DRR synthesis is critical because registration quality depends strongly on consistency between synthesized projections and acquired fluoroscopic images.

Prior studies have investigated the influence of rendering quality, similarity metrics, image preprocessing, and intensity normalization on DRR-based registration performance \cite{penney1998comparison,maier2013precision}. Commonly used similarity measures include normalized cross-correlation, mutual information, structural similarity metrics, and gradient-based measures. These methods primarily evaluate robustness with respect to image noise, anatomical variability, or optimization behavior.

Comparatively little work has examined how calibration perturbations alter vertebral projection appearance despite preserving acceptable reconstruction performance. In particular, the sensitivity of AP and LAT vertebral projections to intrinsic calibration variation remains poorly characterized.

\subsection{Recent Advances in DRR-Based Registration and Differentiable Projection Models}

Recent work has explored differentiable DRR generation, deep-learning-based 2D--3D registration, and uncertainty-aware imaging models for fluoroscopy-guided intervention \cite{gopalakrishnan2022differentiable,miao20163d2d,unberath2017deepdrr}. Differentiable projection frameworks enable gradient-based optimization directly in the image domain, while learning-based methods improve registration robustness under challenging imaging conditions.

Despite these advances, comparatively little attention has been devoted to understanding how intrinsic calibration perturbations influence vertebral projection consistency itself. The current work focuses specifically on projection-domain geometric sensitivity under controlled perturbation conditions.

\subsection{Calibration Ambiguity and Projection Consistency}

Geometric calibration and medical image reconstruction are inherently affected by parameter ambiguity, especially in sparse-view and limited-angle imaging systems. Multiple intrinsic and extrinsic parameter combinations may generate similar reprojection accuracy while producing different local projection geometry.

Most existing calibration validation approaches rely primarily on reconstruction-domain metrics, which may overlook projection-domain inconsistencies relevant to image-guided registration. For vertebral anatomy, small changes in projection geometry can alter pedicle appearance, vertebral contour shape, endplate alignment, and silhouette overlap, potentially degrading registration robustness and increasing target registration error.

This work addresses the gap between reconstruction-domain calibration evaluation and projection-domain registration behavior. We systematically investigate how reconstruction-preserving intrinsic calibration perturbations influence vertebral DRR consistency, AP/LAT projection sensitivity, and downstream 2D--3D registration performance.

\section{Methods}

\subsection{Problem Definition}
The standard pinhole projection model is defined as:

\begin{equation}
\mathbf{x} \sim \mathbf{P}\mathbf{X}
\end{equation}

where $\mathbf{X} \in \mathbb{R}^3$ denotes a 3D point in homogeneous coordinates and $\mathbf{x} \in \mathbb{R}^2$ denotes the corresponding detector projection.

The projection matrix is expressed as:

\begin{equation}
\mathbf{P} = \mathbf{K}[\mathbf{R}|\mathbf{t}]
\end{equation}

where $\mathbf{K}$ denotes the intrinsic calibration matrix and $\mathbf{R}, \mathbf{t}$ denote the extrinsic imaging geometry.

In this work, we investigate how perturbations in intrinsic calibration parameters influence the projected appearance of vertebral anatomy under fixed imaging poses.

Ground-truth imaging geometry is defined as:

\begin{equation}
\mathbf{P}_{gt} = \mathbf{K}_{gt}[\mathbf{R}|\mathbf{t}]
\end{equation}

A perturbed imaging geometry is defined as:

\begin{equation}
\mathbf{P}_{perr} = \mathbf{K}_{perr}[\mathbf{R}|\mathbf{t}]
\end{equation}

where:

\begin{itemize}
    \item $\mathbf{K}_{gt}$ denotes the ground-truth intrinsic calibration matrix,
    \item $\mathbf{K}_{perr}$ denotes the perturbed intrinsic calibration matrix.
\end{itemize}

The extrinsic imaging geometry is held fixed throughout all experiments to isolate the effect of intrinsic calibration perturbations on vertebral projection appearance.

The central hypothesis of this work is:

\begin{quote}
Small perturbations in intrinsic calibration parameters can produce measurable changes in vertebral projection geometry and anatomical shape appearance in AP and LAT views.
\end{quote}

\begin{figure*}[t]
\centering
\includegraphics[width=\textwidth]{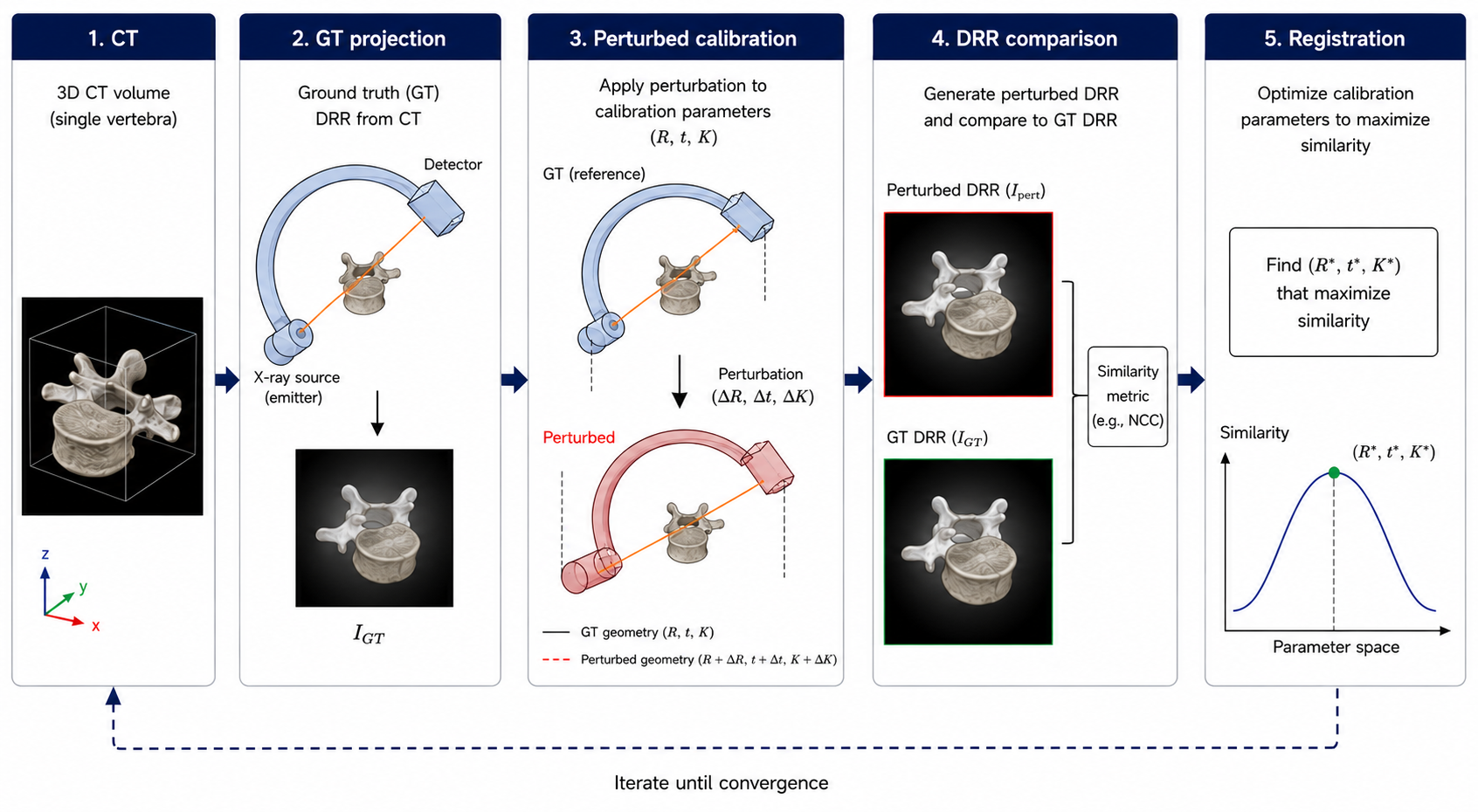}
\caption{
Overview of the proposed projection-domain analysis pipeline. 
A CT-derived vertebral volume is first projected using ground-truth imaging geometry to generate a reference DRR image. 
Controlled perturbations are then applied to the intrinsic calibration parameters to generate perturbed DRRs. 
The perturbed and reference projections are subsequently compared using landmark, contour, silhouette, and image similarity metrics to evaluate projection-domain sensitivity to calibration variation.
}
\label{fig:pipeline}
\end{figure*}

Figure~\ref{fig:pipeline} summarizes the overall calibration perturbation and DRR analysis framework used throughout this study.

\subsection{Synthetic Vertebral Dataset}
Synthetic vertebral datasets are constructed from CT-derived vertebral segmentation masks and mesh models\cite{VanessaTTT-SEG, wang2025hypergraph}. Vertebral geometries are obtained from publicly available datasets including VerSe, SpineWeb, and TotalSegmentator.

Initial experiments are performed using isolated vertebra models to analyze projection sensitivity under controlled conditions. Additional experiments are conducted on multi-vertebra lumbar segments.

Each vertebral mesh is represented as:

\begin{equation}
\mathcal{V} = \{\mathbf{v}_i\}_{i=1}^{N}
\end{equation}

where $\mathbf{v}_i$ denotes a mesh vertex.

Anatomical landmarks including pedicles, vertebral corners, endplates, and spinous processes are annotated for quantitative projection analysis.

\subsection{Synthetic Imaging Geometry}
Synthetic fluoroscopic imaging geometry is constructed using known projection parameters.

The following imaging parameters are defined:

\begin{itemize}
    \item source-to-detector distance (SDD),
    \item source-to-isocenter distance (SID),
    \item detector pixel spacing,
    \item detector dimensions,
    \item AP imaging pose,
    \item LAT imaging pose.
\end{itemize}

Using these parameters, vertebral mesh vertices are projected into the detector plane using:

\begin{equation}
\mathbf{x} \sim \mathbf{P}\mathbf{X}
\end{equation}

Projection contours and landmark locations are generated directly from the projected mesh geometry.

\subsection{Intrinsic Calibration Perturbation Model}

Controlled perturbations were introduced to the intrinsic calibration matrix to simulate uncertainty in C-arm calibration while keeping the extrinsic imaging geometry fixed. This design allowed the effect of intrinsic calibration variation on projection-domain consistency and 2D--3D registration accuracy to be isolated.

The intrinsic calibration matrix was parameterized as:

\begin{equation}
\mathbf{K} =
\begin{bmatrix}
f_x & s & c_x \\
0 & f_y & c_y \\
0 & 0 & 1
\end{bmatrix},
\end{equation}

where $f_x$ and $f_y$ denote the effective focal lengths in detector pixel units, $(c_x,c_y)$ denotes the principal point, and $s$ denotes the detector skew. In the context of C-arm imaging, the principal point is also referred to as the piercing point, which represents the intersection of the central X-ray beam with the detector plane.

In this study, the sensitivity analysis focused on two primary intrinsic calibration perturbations: focal scale and piercing point displacement. The focal scale parameter models uncertainty in the effective source-to-detector magnification or detector pixel scaling by applying a multiplicative factor to the nominal focal lengths. The perturbed focal lengths were defined as:

\begin{equation}
f_x^{perr} = \alpha_f f_x^{gt},
\end{equation}

\begin{equation}
f_y^{perr} = \alpha_f f_y^{gt},
\end{equation}

where $f_x^{gt}$ and $f_y^{gt}$ are the ground-truth focal lengths, $f_x^{perr}$ and $f_y^{perr}$ are the perturbed focal lengths, and $\alpha_f$ is the focal scale factor. A value of $\alpha_f = 1$ corresponds to the ground-truth calibration. Values of $\alpha_f > 1$ and $\alpha_f < 1$ simulate overestimation and underestimation of the effective focal length, respectively.

Piercing point displacement models uncertainty in the assumed location where the central ray intersects the detector. Since the piercing point corresponds to the principal point in the pinhole projection model, this perturbation was represented as a 2D shift of $(c_x,c_y)$ on the detector plane:

\begin{equation}
c_x^{perr} = c_x^{gt} + \Delta c_x,
\end{equation}

\begin{equation}
c_y^{perr} = c_y^{gt} + \Delta c_y,
\end{equation}

where $(c_x^{gt},c_y^{gt})$ denotes the ground-truth piercing point, $(c_x^{perr},c_y^{perr})$ denotes the perturbed piercing point, and $(\Delta c_x,\Delta c_y)$ denotes the detector-plane displacement. The displacement was expressed in pixels unless otherwise specified. Positive and negative values of $\Delta c_x$ and $\Delta c_y$ correspond to horizontal and vertical shifts of the assumed piercing point on the detector.

Using these definitions, the perturbed intrinsic calibration matrix was written as:

\begin{equation}
\mathbf{K}_{perr} =
\begin{bmatrix}
\alpha_f f_x^{gt} & s^{gt} & c_x^{gt} + \Delta c_x \\
0 & \alpha_f f_y^{gt} & c_y^{gt} + \Delta c_y \\
0 & 0 & 1
\end{bmatrix}.
\end{equation}

Equivalently, the intrinsic perturbation can be written in additive form as:

\begin{equation}
\mathbf{K}*{perr} = \mathbf{K}*{gt} + \Delta \mathbf{K},
\end{equation}

where $\Delta \mathbf{K}$ represents the controlled perturbation applied to the intrinsic calibration parameters. In this work, focal scale changes the effective projection magnification, whereas piercing point displacement shifts the assumed detector location of the central ray. These two perturbations were applied independently in the sensitivity analysis to isolate their respective effects.

The extrinsic imaging geometry was kept fixed for all perturbation experiments:

\begin{equation}
\mathbf{R}*{err} = \mathbf{R}*{gt},
\end{equation}

\begin{equation}
\mathbf{t}*{err} = \mathbf{t}*{gt}.
\end{equation}

Thus, any observed changes in projected landmark location, contour geometry, silhouette overlap, or registration accuracy were attributed to intrinsic calibration perturbation rather than changes in the acquisition pose.

\subsection{Projection-Domain Analysis}

Projection consistency is evaluated by comparing vertebral projections generated using ground-truth and perturbed intrinsic calibration parameters.

\subsubsection{Landmark Projection Error}

Projection displacement of anatomical landmarks is computed as:

\begin{equation}
e_{2D} = ||\mathbf{x}_{gt} - \mathbf{x}_{err}||_2
\end{equation}

where $\mathbf{x}_{gt}$ and $\mathbf{x}_{err}$ denote landmark projections under ground-truth and perturbed calibration, respectively.

The following anatomical structures are evaluated:

\begin{itemize}
    \item pedicles,
    \item vertebral corners,
    \item spinous processes,
    \item superior and inferior endplates.
\end{itemize}

\subsubsection{Contour Shape Analysis}

Projection contour consistency is quantified using:

\begin{itemize}
    \item Chamfer distance,
    \item Hausdorff distance,
    \item edge-based contour distance.
\end{itemize}

Let $\mathcal{C}_{gt}$ and $\mathcal{C}_{err}$ denote the projected contours generated using ground-truth and perturbed calibration parameters.

The Chamfer distance is computed as:

\begin{equation}
D_{Chamfer}(\mathcal{C}_{gt}, \mathcal{C}_{err})
=
\sum_{x \in \mathcal{C}_{gt}}
\min_{y \in \mathcal{C}_{err}}
||x-y||_2
\end{equation}

\subsubsection{Silhouette Overlap}

Projection silhouette consistency is evaluated using Dice similarity coefficient and intersection-over-union (IoU).

The Dice coefficient is defined as:

\begin{equation}
Dice =
\frac{2|A \cap B|}
{|A| + |B|}
\end{equation}

where $A$ and $B$ denote binary vertebral projection masks.

\subsection{Landmark-Based 2D--3D Registration}

To evaluate the downstream effect of intrinsic calibration perturbations on registration accuracy, a landmark-based 2D--3D registration experiment was performed. The objective was to estimate the rigid transformation that aligns a 3D vertebral model to its corresponding 2D fluoroscopic projections under different intrinsic calibration conditions.

Let $\mathbf{X}*i \in \mathbb{R}^3$ denote the $i$-th anatomical landmark defined on the vertebral model, and let $\mathbf{x}*{i}^{gt} \in \mathbb{R}^2$ denote its corresponding ground-truth 2D projection generated using the reference imaging geometry. The set of 3D landmarks included vertebral body corners, pedicle centers, superior and inferior endplate points, and the spinous process tip. These landmarks were selected because they are anatomically meaningful and can be consistently identified on vertebral projections.

The vertebral pose was represented by a rigid transformation $\mathbf{T} \in SE(3)$:

\begin{equation}
  \mathbf{T} =
  \begin{bmatrix}
    \mathbf{R}_{\mathrm{reg}} & \mathbf{t}_{\mathrm{reg}} \\
    \mathbf{0}_{1 \times 3} & 1
  \end{bmatrix}.
\end{equation}

where $\mathbf{R}*{reg}$ and $\mathbf{t}*{reg}$ denote the estimated registration rotation and translation, respectively. During registration, the transformed 3D landmark $\mathbf{T}\mathbf{X}_i$ was projected into the detector plane using the assumed projection matrix $\mathbf{P}$:

\begin{equation}
\hat{\mathbf{x}}_i(\mathbf{T}) \sim \mathbf{P}\mathbf{T}\mathbf{X}_i .
\end{equation}

The landmark-based registration objective was defined as the minimization of the 2D reprojection error between the observed landmark locations and the projected 3D landmark locations:

\begin{equation}
\mathbf{T}^{*}=\arg\min_{\mathbf{T} \in SE(3)}\sum_{i=1}^{N}\left|\mathbf{x}_{i}^{gt}-\hat{\mathbf{x}}_i(\mathbf{T})\right|_2^2 .
\end{equation}

For single-view registration, the objective function was computed using either the AP or LAT projection independently. For multi-view registration, AP and LAT projection errors were jointly minimized:

\begin{equation}
\mathbf{T}^{*}=\arg\min_{\mathbf{T} \in SE(3)}\sum_{v \in {AP,LAT}}\sum_{i=1}^{N}\left|\mathbf{x}_{i,v}^{gt}-\hat{\mathbf{x}}_{i,v}(\mathbf{T})\right|_2^2 ,
\end{equation}

where $v$ denotes the projection view. This formulation allowed registration performance to be compared between AP-only, LAT-only, and combined AP/LAT configurations.

Registration was initialized from the ground-truth vertebral pose with controlled perturbations applied to the initial rotation and translation. The six rigid transformation parameters, consisting of three rotational parameters and three translational parameters, were optimized iteratively. Rotation was parameterized using Euler angles or an equivalent axis-angle representation, and translation was parameterized in millimeters along the three anatomical axes.

Optimization was performed using a nonlinear least-squares procedure. At each iteration, the current 3D vertebral landmarks were transformed by the estimated pose, projected into the detector plane using the assumed calibration matrix, and compared with the reference 2D landmark locations. The pose parameters were then updated to reduce the total reprojection error.

The convergence criterion was defined based on both the change in objective function value and the change in estimated pose parameters. Optimization was terminated when the relative decrease in reprojection error was below a predefined threshold, when the pose update magnitude became smaller than a predefined tolerance, or when the maximum number of iterations was reached.

To evaluate the effect of calibration perturbation, registration was performed using projection matrices constructed from both the ground-truth intrinsic matrix $\mathbf{K}*{gt}$ and the perturbed intrinsic matrix $\mathbf{K}*{perr}$. The resulting registration estimates were compared against the known ground-truth vertebral pose. Registration accuracy was quantified using translational error, rotational error, and residual 2D reprojection error.

The translational registration error was computed as:

\begin{equation}
e_t =
\left|
\mathbf{t}_{reg}-\mathbf{t}_{gt}\right|_2 ,
\end{equation}

where $\mathbf{t}*{reg}$ and $\mathbf{t}*{gt}$ denote the estimated and ground-truth translations, respectively.

The rotational registration error was computed from the relative rotation matrix:

\begin{equation}
\Delta \mathbf{R}=\mathbf{R}*{reg} - \mathbf{R}*{gt}^{T},
\end{equation}

and expressed as the geodesic angular error:

\begin{equation}
e_R =
\cos^{-1}
\left(
\frac{
\mathrm{trace}(\Delta \mathbf{R}) - 1
}{2}
\right).
\end{equation}

Residual reprojection error was computed after convergence as:

\begin{equation}
e_{rep}=\frac{1}{N}\sum_{i=1}^{N}\left|\mathbf{x}_{i}^{gt}-\hat{\mathbf{x}}_i(\mathbf{T}^{*})\right|_2 .
\end{equation}

For multi-view registration, the residual reprojection error was averaged across all landmarks and all projection views. These metrics were computed separately for AP-only, LAT-only, and AP/LAT registration conditions to determine whether calibration perturbations produced view-dependent degradation in 2D--3D registration accuracy.

\subsection{AP/LAT Sensitivity Analysis}

Projection sensitivity is analyzed independently for AP and LAT imaging geometries.

We hypothesize that LAT projections exhibit greater sensitivity to intrinsic calibration perturbations due to:

\begin{itemize}
    \item increased depth compression,
    \item overlapping anatomical structures,
    \item reduced geometric separation between vertebral features.
\end{itemize}

Projection sensitivity is quantitatively compared between AP and LAT views using landmark displacement, contour mismatch, and silhouette overlap metrics.

\subsection{Anatomical Sensitivity Analysis}

Sensitivity analysis is additionally performed for individual vertebral structures including:

\begin{itemize}
    \item vertebral body,
    \item pedicles,
    \item spinous process,
    \item superior and inferior endplates.
\end{itemize}

This analysis identifies which anatomical structures exhibit the greatest projection sensitivity to intrinsic calibration perturbations.

\subsection{Experimental Configuration}

Experiments are performed using CT-derived lumbar vertebrae including vertebral levels L1--L5. Both isolated single-vertebra models and multi-vertebra lumbar segments are evaluated. CT volumes are resampled to isotropic voxel spacing prior to mesh extraction and DRR generation.

Synthetic DRRs are generated at a detector resolution of $1024 \times 1024$ pixels using a cone-beam projection geometry. The detector pixel spacing, detector dimensions, source-to-detector distance (SDD), and source-to-isocenter distance (SID) are fixed throughout all experiments to ensure consistent imaging geometry.

Two standard fluoroscopic imaging configurations are evaluated:

\begin{itemize}
    \item anterior--posterior (AP) view,
    \item lateral (LAT) view.
\end{itemize}

For each vertebral model, intrinsic calibration perturbations are independently applied to focal length, principal point location, detector skew, and pixel spacing parameters. Perturbation magnitudes are sampled across predefined ranges to simulate varying levels of calibration uncertainty.

Multiple perturbation trials are performed for each parameter configuration to evaluate projection consistency under repeated calibration variation conditions.

Anatomical landmarks are manually or semi-automatically defined using vertebral anatomical reference points, including pedicle centers, vertebral body corners, spinous process tips, and superior/inferior endplate boundaries.

DRR generation is implemented using ray-driven projection rendering based on line-integral attenuation accumulation through CT volumes. Projection contour extraction and silhouette generation are computed directly from the rendered vertebral masks.

All experiments and analyses are implemented in Python using NumPy, SciPy, OpenCV, and PyTorch-based image processing utilities. Quantitative metrics are computed independently for AP and LAT views for all perturbation conditions.

\section{Experiments}

\subsection{Experimental Setup and Imaging Geometry}
\label{sec:experimental_setup}

Experiments were conducted using synthetic vertebral projection data generated from CT-derived vertebral mesh models under controlled fluoroscopic imaging geometry. Two standard projection views were simulated: anteroposterior (AP) and lateral (LAT). For both views, the vertebral model was kept fixed in the same anatomical pose, and no inter-view motion, repositioning, or deformation was introduced. Thus, differences observed between AP and LAT projections were not caused by anatomical motion, but instead reflected the combined influence of view-dependent imaging geometry and intrinsic calibration perturbation.

\begin{figure}[t]
    \centering
    \includegraphics[width=0.85\textwidth]{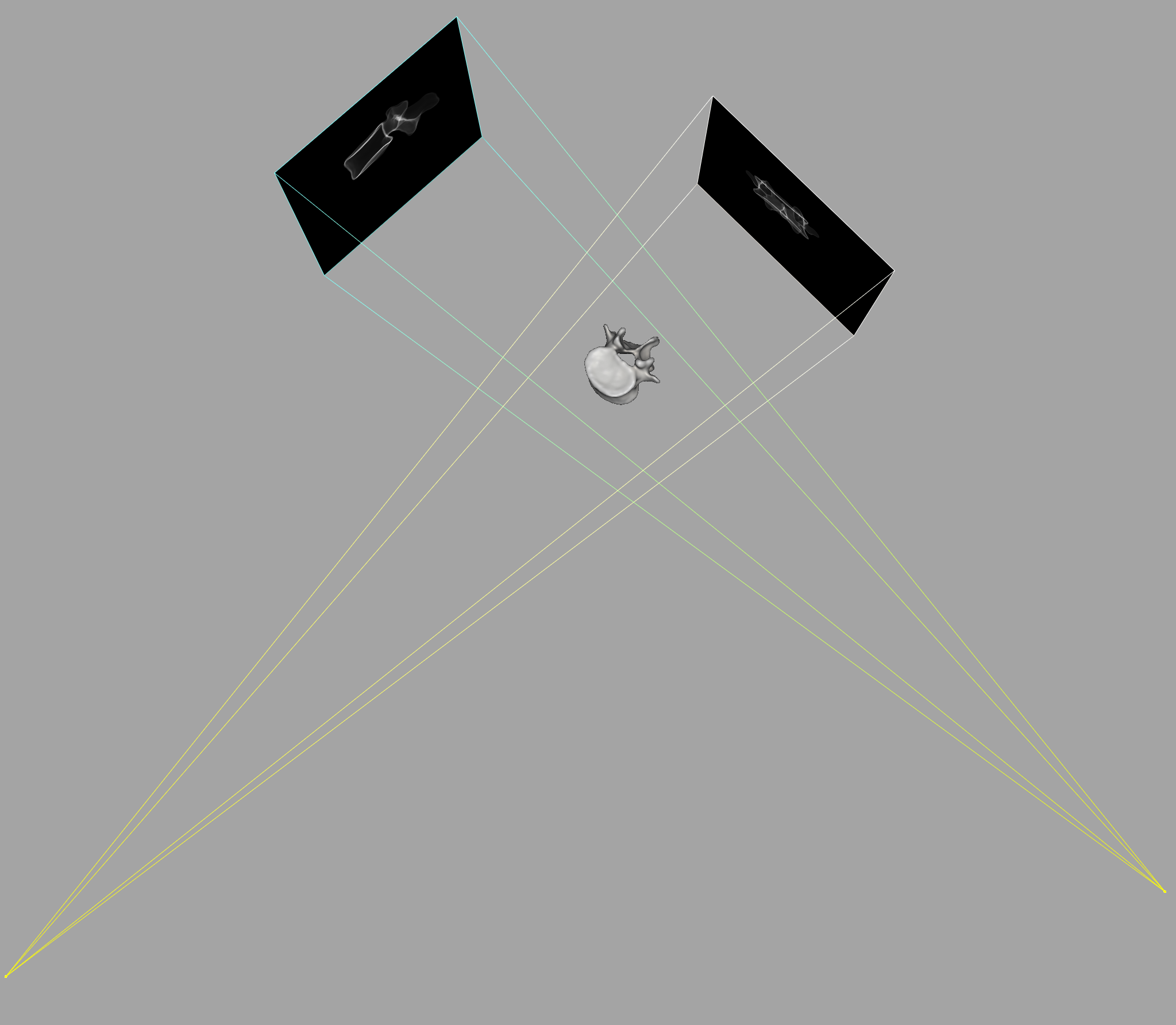}
    \caption{
    Schematic illustration of the simulated AP and LAT fluoroscopic imaging setup. 
    The vertebral model is assumed to remain fixed between views, so no inter-view anatomical motion, repositioning, or deformation is introduced. 
    The modeled C-arm geometry is non-isocentric; therefore, the source--object--detector relationship differs between AP and LAT projections. 
    In the LAT configuration, the vertebra is closer to the detector and appears larger in the projected image, contributing to increased projection-domain sensitivity to intrinsic calibration perturbation.
    }
    \label{fig:imaging_setup}
\end{figure}

The imaging setup is illustrated in Figure~\ref{fig:imaging_setup}. The modeled C-arm geometry was non-isocentric, meaning that the source--object--detector relationship was not identical between AP and LAT views. In particular, in the LAT configuration, the vertebra was positioned closer to the detector and therefore appeared larger in the projected image. This difference in object--detector geometry is important for interpreting the results, because the same intrinsic calibration perturbation can produce larger apparent landmark displacement and contour deformation when the projected anatomy occupies a larger region of the detector.

Ground-truth AP and LAT projections were first generated using known intrinsic and extrinsic imaging parameters. Intrinsic calibration perturbations were then introduced while keeping the vertebral anatomy and the view-specific extrinsic geometry fixed. In other words, for each view, the ground-truth and perturbed projections differed only in the intrinsic calibration matrix. This design allowed projection-domain deformation to be attributed specifically to intrinsic calibration variation rather than to changes in acquisition pose or vertebral position.

The primary experiments focused on focal length perturbation, which produced the largest projection-domain sensitivity. The ground-truth focal length was approximately 4500 pixels, and controlled focal length perturbations ranging from 100 to 500 pixels were applied to evaluate sensitivity under increasing calibration error magnitude. Additional intrinsic perturbations, including principal point displacement, detector pixel spacing variation, and detector skew, were also evaluated. Principal point perturbations ranging from 10 to 50 pixels were used in the corresponding experiments.

Initial experiments were performed using isolated vertebral models to enable controlled analysis of projection sensitivity. Anatomical landmarks, including the vertebral center, pedicles, superior and inferior endplates, and spinous process tip, were manually annotated for quantitative evaluation.

\subsection{Intrinsic Perturbation Study}
Intrinsic calibration perturbations were incrementally increased to analyze their effect on vertebral projection morphology.

For each perturbation configuration, vertebral projections generated using perturbed intrinsic parameters were compared against ground-truth projections generated using the original calibration matrix.

Projection sensitivity was quantitatively evaluated using both global silhouette consistency metrics and local anatomical displacement measurements.

Global projection consistency was evaluated using:

\begin{itemize}
    \item Dice similarity coefficient,
    \item intersection-over-union (IoU),
    \item contour overlap analysis.
\end{itemize}

Local anatomical sensitivity was evaluated using:

\begin{itemize}
    \item landmark projection displacement,
    \item contour deformation,
    \item anatomical structure shift.
\end{itemize}

This experiment established the relationship between intrinsic calibration perturbation magnitude and projection-domain deformation.

\subsection{AP/LAT Projection Sensitivity Analysis}
Projection sensitivity was analyzed independently for AP and LAT imaging geometries.

For each imaging view, vertebral projections generated using perturbed intrinsic parameters were compared against ground-truth projections using both qualitative and quantitative analysis.

Quantitative evaluation included:

\begin{itemize}
    \item landmark reprojection displacement,
    \item Dice similarity coefficient,
    \item intersection-over-union (IoU),
    \item contour deformation analysis.
\end{itemize}

Qualitative overlay visualizations were additionally generated to directly compare contour deformation between AP and LAT projections.

The primary objective of this analysis was to evaluate whether identical intrinsic calibration perturbations produce different projection-domain effects across imaging orientations.

\subsection{Anatomical Structure Sensitivity Analysis}
Projection sensitivity was additionally analyzed for individual vertebral anatomical structures including:

\begin{itemize}
    \item vertebral center,
    \item superior endplate,
    \item inferior endplate,
    \item left and right pedicles,
    \item spinous process tip.
\end{itemize}

For each anatomical structure, landmark displacement was measured between ground-truth and perturbed projections.

This analysis enabled evaluation of which vertebral structures exhibited the greatest sensitivity to intrinsic calibration perturbation and whether anatomical sensitivity differed between AP and LAT imaging geometries.

\subsection{2D--3D Registration Error Evaluation}

To further evaluate the influence of intrinsic calibration perturbation on downstream registration behavior, synthetic 2D--3D registration experiments were performed using vertebral projections generated under varying intrinsic calibration conditions. Registration error was evaluated across multiple focal scaling perturbations and simulated anatomical piercing conditions.

Intrinsic calibration perturbations were introduced by modifying focal scaling parameters while preserving the underlying vertebral anatomy and imaging pose. Focal scale values ranged from 100 to 500, and piercing conditions were varied from 0 to 50 in incremental steps. For each perturbation condition, vertebral projections were generated and aligned using two registration strategies:

\begin{enumerate}
    \item \textbf{Kabsch-based rigid registration}, which estimates optimal rigid alignment using corresponding landmark points.
    \item \textbf{Direct registration}, which directly evaluates projection alignment under perturbed calibration conditions.
\end{enumerate}

Registration performance was quantified using rotational error (degrees) and translational error (mm)\cite{ZHOU2021104923}. Rotation error was computed from the angular deviation between estimated and reference rigid transformations, while translation error was measured as the Euclidean displacement between corresponding translation vectors.

The experiments were designed to isolate the effects of intrinsic calibration perturbation on projection-domain registration behavior while maintaining fixed vertebral anatomy and acquisition geometry.

\section{Results}

\subsection{Qualitative Projection-Domain Difference Analysis}

Qualitative projection-domain analysis was performed to visualize how intrinsic focal length perturbations affect vertebral DRR appearance in AP and LAT imaging geometries. Instead of using binary projection overlays, we computed raw DRR difference heatmaps between the reference DRR generated with the ground-truth intrinsic calibration and the perturbed DRR generated after modifying the focal length. Each heatmap represents the absolute pixel-wise raw intensity difference, thereby highlighting local projection-domain discrepancies caused by calibration variation.

\begin{figure}[t]
    \centering
    
    \begin{subfigure}[b]{0.48\textwidth}
        \centering
        \includegraphics[width=\textwidth]{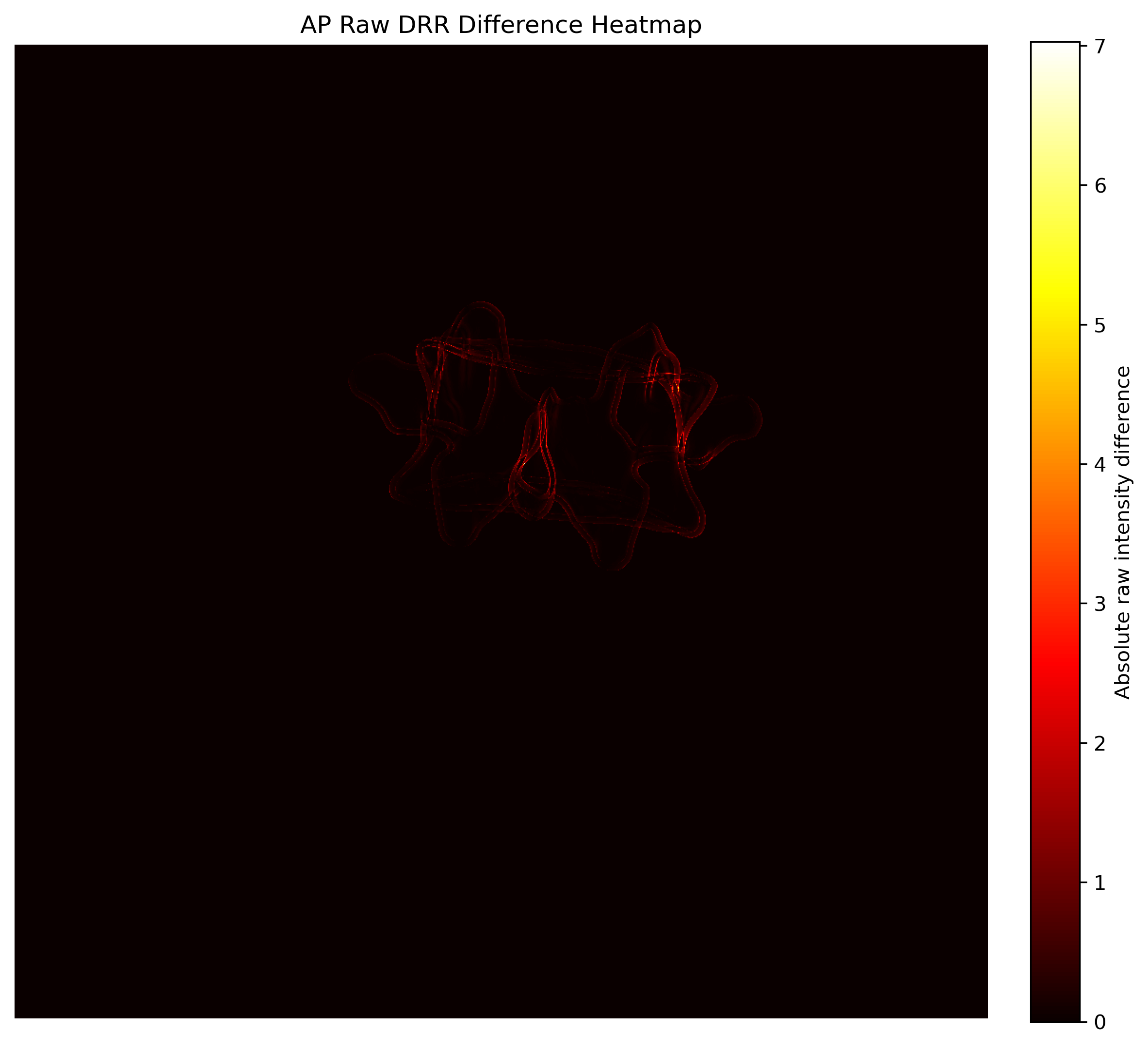}
        \caption{AP, 100-pixel focal length perturbation.}
        \label{fig:ap_raw_difference_heatmap_100}
    \end{subfigure}
    \hfill
    \begin{subfigure}[b]{0.48\textwidth}
        \centering
        \includegraphics[width=\textwidth]{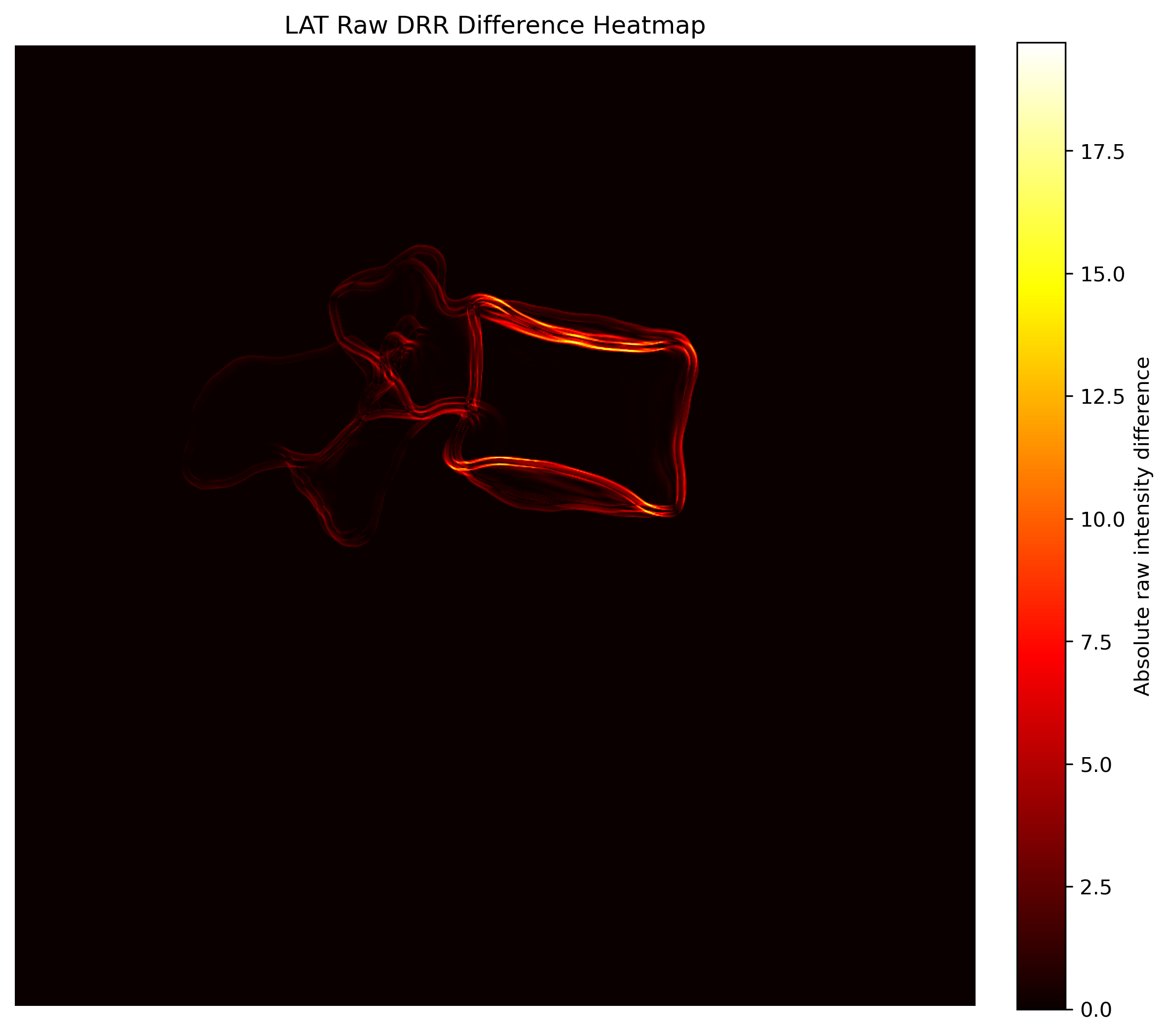}
        \caption{LAT, 100-pixel focal length perturbation.}
        \label{fig:lat_raw_difference_heatmap_100}
    \end{subfigure}

    \vspace{0.5em}

    \begin{subfigure}[b]{0.48\textwidth}
        \centering
        \includegraphics[width=\textwidth]{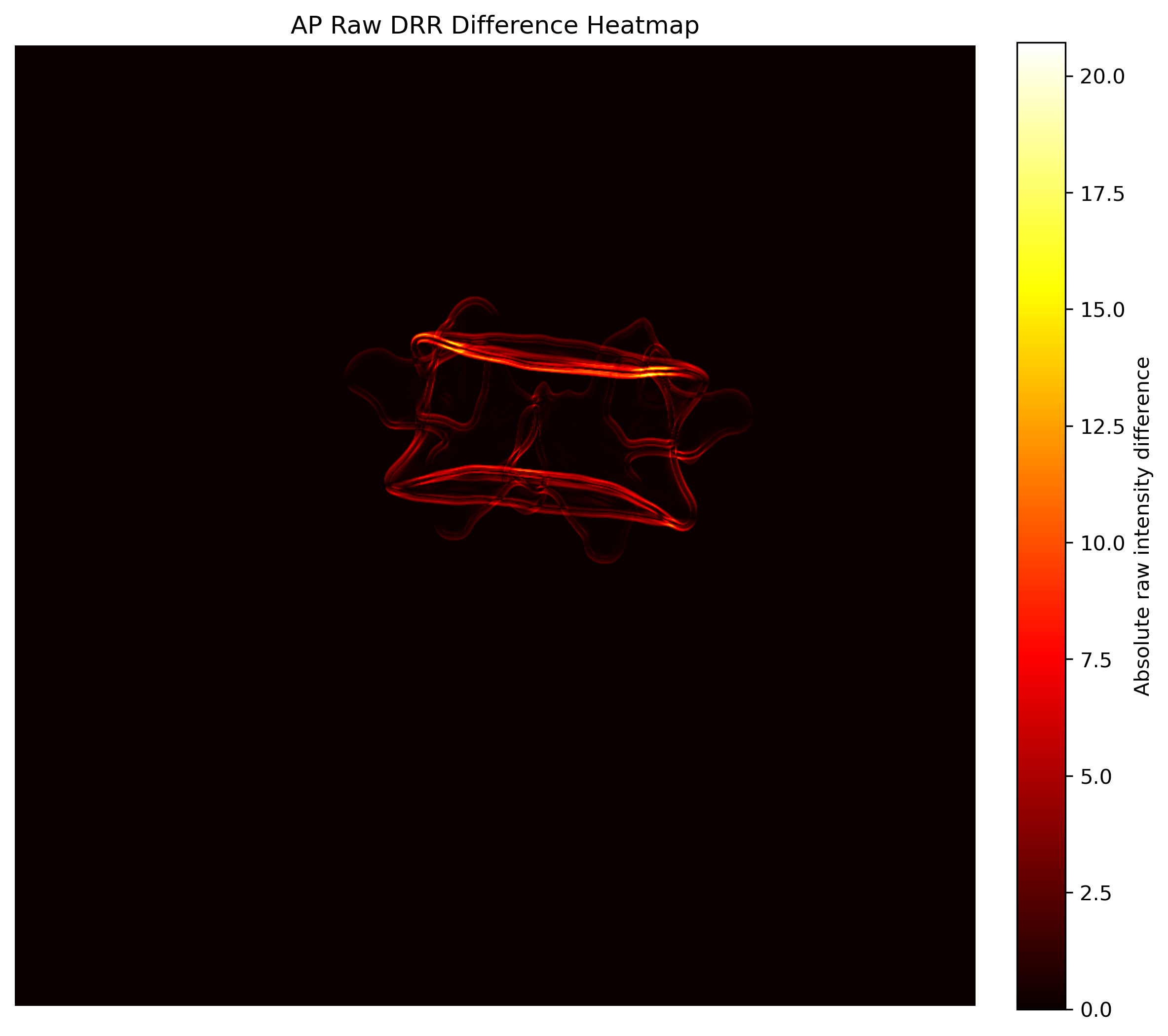}
        \caption{AP, 500-pixel focal length perturbation.}
        \label{fig:ap_raw_difference_heatmap_500}
    \end{subfigure}
    \hfill
    \begin{subfigure}[b]{0.48\textwidth}
        \centering
        \includegraphics[width=\textwidth]{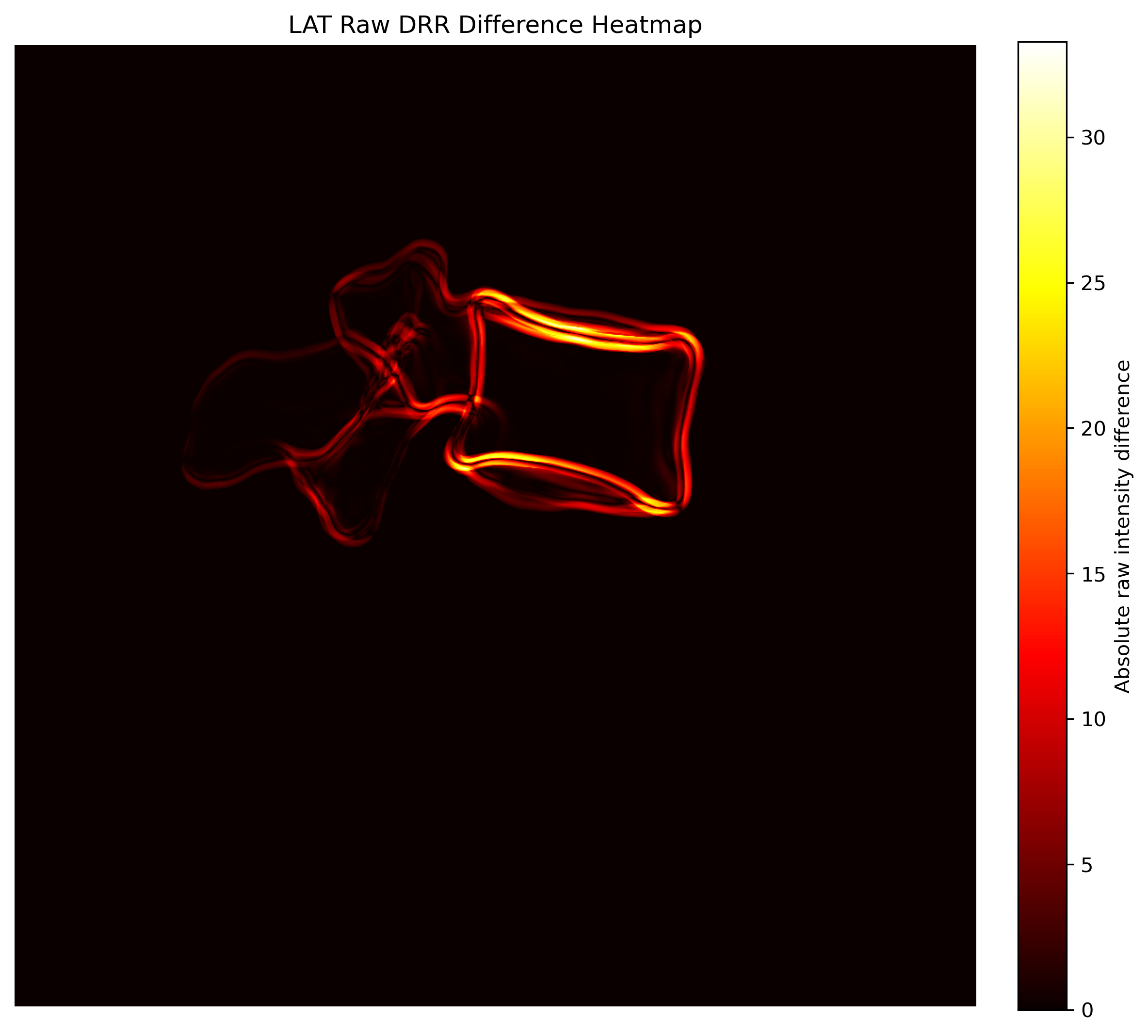}
        \caption{LAT, 500-pixel focal length perturbation.}
        \label{fig:lat_raw_difference_heatmap_500}
    \end{subfigure}

    \caption{AP and LAT raw DRR difference heatmaps under focal length perturbations of 100 and 500 pixels. The heatmaps show the absolute pixel-wise raw intensity difference between reference and perturbed DRRs. Increasing the perturbation magnitude increases the difference response in both views, while the LAT projection consistently exhibits stronger and more spatially extended discrepancies than the AP projection.}
    
    \label{fig:raw_difference_heatmap_multiscale}
\end{figure}

Figure~\ref{fig:raw_difference_heatmap_multiscale} shows AP and LAT raw DRR difference heatmaps under focal length perturbations of 100 and 500 pixels. For both perturbation magnitudes, the difference responses are concentrated around high-gradient anatomical structures, including vertebral body boundaries, cortical margins, pedicle regions, and posterior elements. As expected, increasing the focal length perturbation from 100 to 500 pixels increases the magnitude and spatial extent of the raw intensity differences in both views.

However, the response is strongly view-dependent. In the AP projection, the difference signal remains relatively weak and spatially localized, even under the larger 500-pixel perturbation. The visible differences are mainly confined to thin boundary regions and do not substantially alter the overall vertebral projection appearance. In contrast, the LAT projection exhibits stronger and more spatially extended differences under the same perturbation magnitudes. Prominent intensity changes are observed along the vertebral body margins, endplate-like structures, pedicle region, and posterior anatomical contours.

These observations indicate that intrinsic focal length perturbation affects AP and LAT projections differently. Although both views are generated from the same vertebral anatomy and are subjected to the same intrinsic perturbation magnitude, the LAT view shows greater projection-domain sensitivity. This increased sensitivity is likely related to the lateral imaging geometry, where depth-dependent magnification changes and overlapping anatomical structures produce larger apparent changes in projection appearance. Overall, the raw DRR difference heatmaps demonstrate that LAT projections are more susceptible than AP projections to focal length calibration variation, even when the perturbation magnitude is reduced from 500 pixels to 100 pixels.

\subsection{Projection Sensitivity versus Intrinsic Perturbation Magnitude}

To quantitatively evaluate projection sensitivity, Dice similarity coefficient and intersection-over-union (IoU) were measured under progressively increasing intrinsic calibration perturbation magnitudes.

\begin{figure}[t]
    \centering
    
    \begin{subfigure}[b]{0.48\textwidth}
        \centering
        \includegraphics[width=\textwidth]{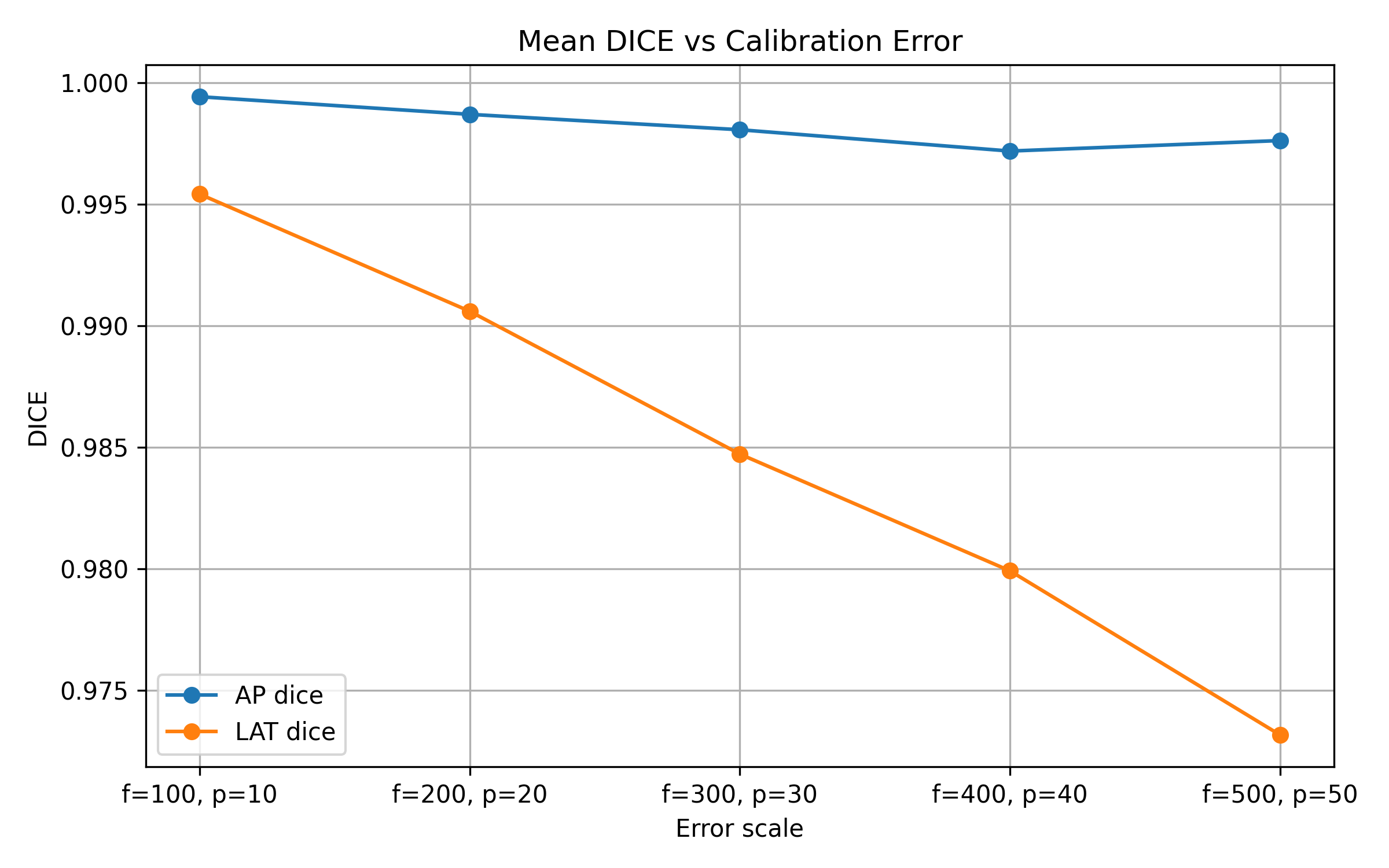}
        \caption{Mean Dice similarity coefficient versus intrinsic calibration perturbation magnitude. LAT projections demonstrate progressively decreasing silhouette similarity under increasing perturbation, whereas AP projections remain comparatively stable.}
        \label{fig:dice_curve}
    \end{subfigure}
    \hfill
    \begin{subfigure}[b]{0.48\textwidth}
        \centering
        \includegraphics[width=\textwidth]{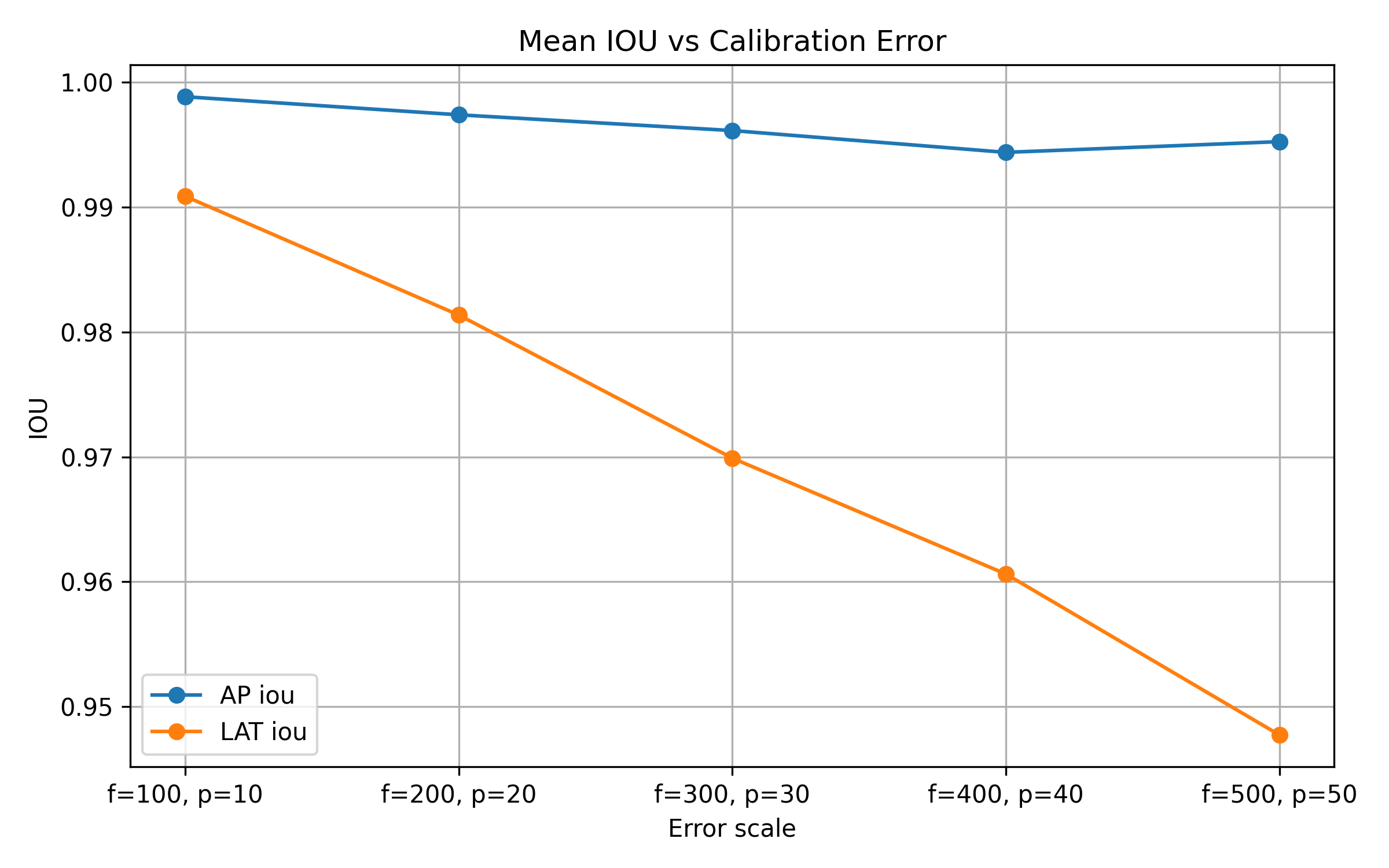}
        \caption{Mean IoU versus intrinsic calibration perturbation magnitude. LAT projections exhibit substantially greater sensitivity to intrinsic perturbation compared to AP projections.}
        \label{fig:iou_curve}
    \end{subfigure}

    \caption{Projection silhouette consistency under increasing intrinsic calibration perturbation magnitude. AP projections remain relatively stable across perturbation scales, while LAT projections demonstrate progressively increasing global shape deformation and silhouette inconsistency.}
    
    \label{fig:dice_iou_curve}
\end{figure}

Figure~\ref{fig:dice_iou_curve} illustrates the relationship between intrinsic perturbation scale and projection silhouette consistency for both AP and LAT imaging geometries. Perturbation magnitude was jointly increased using focal length perturbation ($f$) and principal point displacement ($p$).

As perturbation magnitude increased, LAT projections demonstrated substantially larger degradation in silhouette consistency compared to AP projections. In particular, the LAT Dice coefficient decreased from approximately 0.995 to 0.973 as focal length perturbation increased from $f=100$ pixels to $f=500$ pixels. Similarly, LAT IoU decreased from approximately 0.991 to 0.948 across the same perturbation range.

In contrast, AP projections remained comparatively stable under increasing intrinsic perturbation. AP Dice values remained above 0.997 across all perturbation scales, while AP IoU remained near 0.995 with only minimal variation.

These results demonstrate a clear relationship between intrinsic calibration perturbation magnitude and projection consistency. Although identical intrinsic perturbations were applied to both imaging views, LAT projections exhibited substantially greater global shape deformation and silhouette inconsistency compared to AP projections.

Importantly, even relatively high Dice and IoU values in LAT projections were associated with visible local anatomical displacement, indicating that overlap-based metrics alone may not fully capture localized projection deformation.

\subsection{Landmark Projection Displacement}

\subsection{Landmark Projection Displacement}

Quantitative landmark analysis demonstrates substantial view-dependent sensitivity to intrinsic calibration perturbation.

\begin{figure}[t]
    \centering
    
    \begin{subfigure}[b]{0.48\textwidth}
        \centering
        \includegraphics[width=\textwidth]{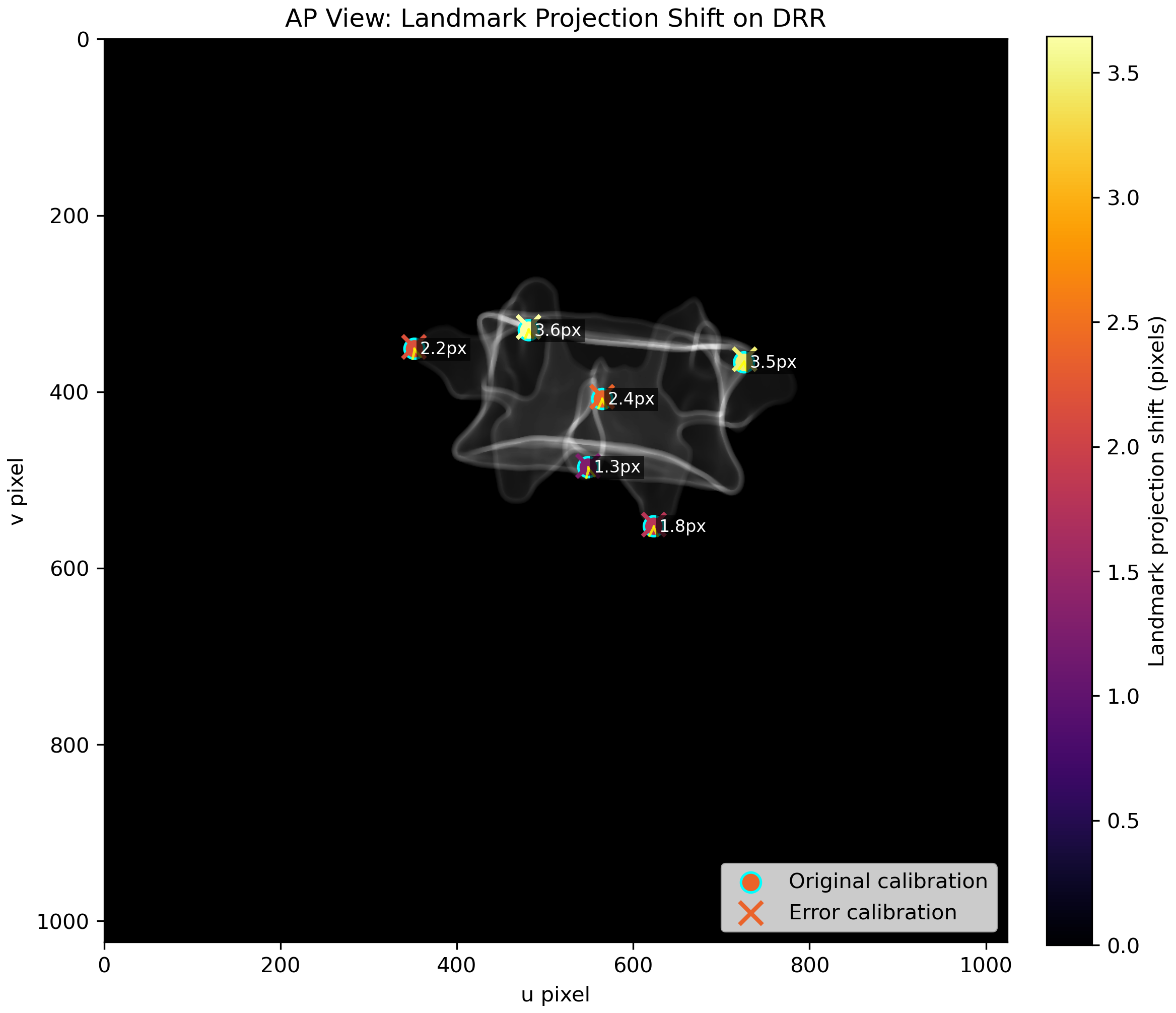}
        \caption{AP landmark displacement under intrinsic focal length perturbation. All anatomical landmarks remain below approximately 2 pixels displacement.}
        \label{fig:ap_landmark_shift}
    \end{subfigure}
    \hfill
    \begin{subfigure}[b]{0.48\textwidth}
        \centering
        \includegraphics[width=\textwidth]{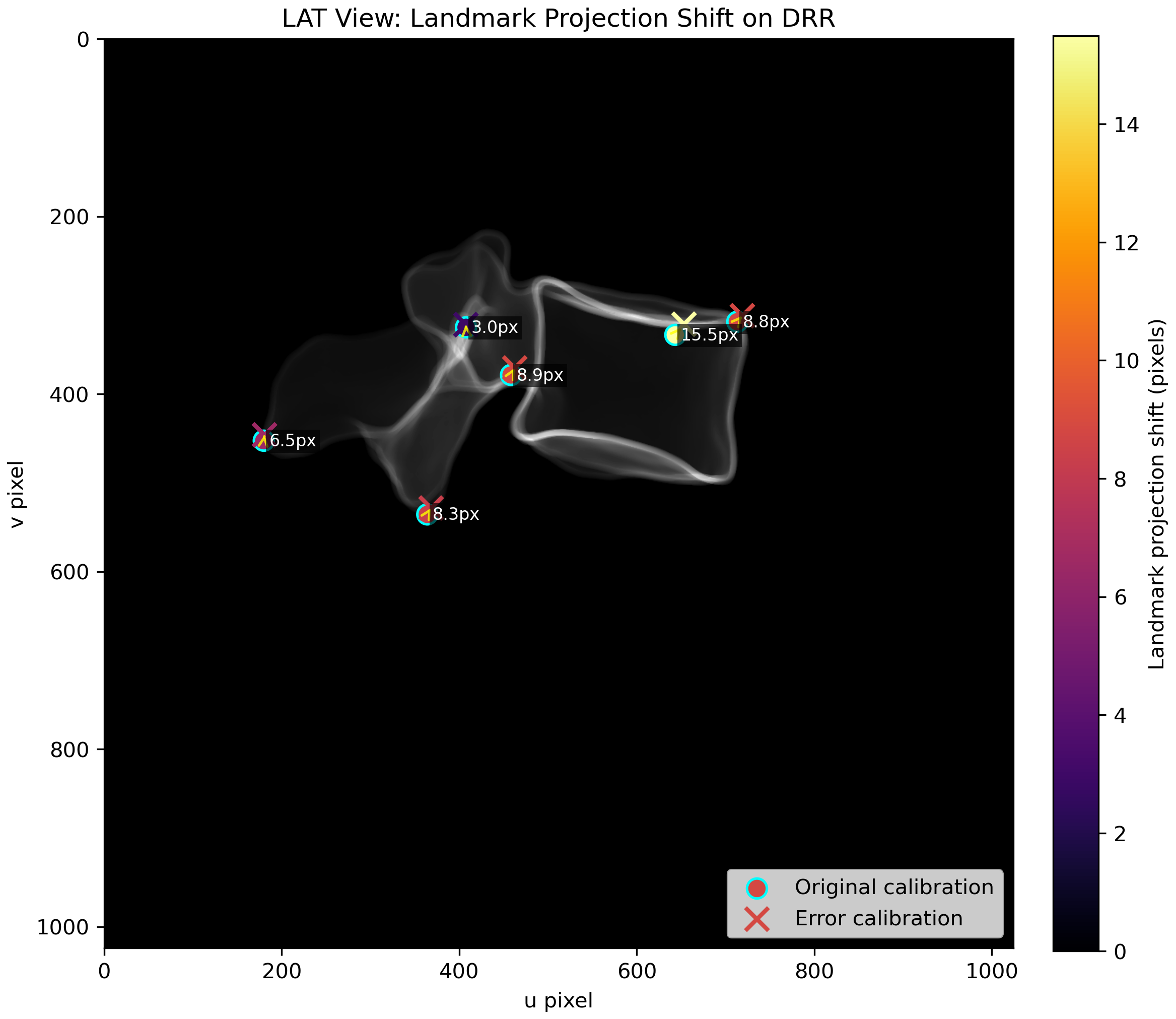}
        \caption{LAT landmark displacement under the same focal length perturbation. Multiple anatomical structures exhibit displacement exceeding 10 pixels.}
        \label{fig:lat_landmark_shift}
    \end{subfigure}

    \caption{Comparison of vertebral landmark displacement between AP and LAT projections under intrinsic calibration perturbation. Blue markers denote ground-truth landmark locations and orange markers denote projections generated using perturbed intrinsic parameters. LAT projections demonstrate substantially greater anatomical displacement compared to AP projections under identical perturbation magnitude.}
    
    \label{fig:landmark_shift}
\end{figure}

Figure~\ref{fig:ap_landmark_shift} and Figure~\ref{fig:lat_landmark_shift} compare anatomical landmark displacement between ground-truth and perturbed intrinsic calibration projections for both AP and LAT imaging geometries. 
The same focal length perturbation was applied in both views, and the vertebral pose was kept fixed between AP and LAT projections. 
Therefore, the measured displacement reflects the combined effect of calibration perturbation and view-dependent imaging geometry, rather than inter-view anatomical motion.

A larger displacement was observed in the LAT projection. 
This behavior is consistent with the non-isocentric C-arm setup, in which the vertebra is closer to the detector in the LAT view and therefore appears larger in the projected image. 
Under this geometry, intrinsic calibration inaccuracies have a more pronounced effect on the projection location of anatomical landmarks.

In the LAT view, large anatomical shifts were observed across multiple vertebral structures. 
The inferior endplate and left pedicle demonstrated the largest displacement, with projection shifts of approximately 21.79 pixels and 21.92 pixels, respectively. 
The vertebral center, spinous tip, and superior endplate also exhibited substantial displacement, ranging from approximately 9.95 to 15.53 pixels.

In contrast, AP projections remained comparatively stable under the same perturbation magnitude. 
All evaluated anatomical landmarks exhibited displacement below 2 pixels, with most landmarks remaining below 1 pixel.

These findings indicate that identical intrinsic calibration perturbations can produce substantially different projection-domain effects depending on imaging orientation and object-detector geometry. 
In the present non-isocentric setup, the LAT view was more sensitive to intrinsic calibration inaccuracies because the vertebra was projected at a larger apparent scale.

Table~\ref{tab:landmark_shift} summarizes the measured landmark displacement for all evaluated anatomical structures.

\begin{table}[t]
\centering
\caption{Comparison of anatomical landmark displacement between AP and LAT projections under intrinsic focal length perturbation.}
\label{tab:landmark_shift}
\begin{tabular}{lcc}
\hline
\textbf{Landmark} & \textbf{AP Shift (px)} & \textbf{LAT Shift (px)} \\
\hline
Center & 2.2 & 6.5 \\
Superior endplate & 3.6 & 3.0 \\
Inferior endplate & 3.5 & 15.5 \\
Left pedicle & 2.4 & 8.9 \\
Right pedicle & 1.3 & 18.3 \\
Spinous tip & 1.8 & 18.8 \\
\hline
\end{tabular}
\end{table}

\subsection{AP versus LAT Sensitivity}

LAT projections demonstrated substantially greater sensitivity to intrinsic calibration perturbations compared to AP projections.

Under identical focal length perturbation, LAT views exhibited larger contour deformation, increased landmark reprojection displacement, and reduced silhouette overlap consistency. In contrast, AP projections remained comparatively stable under the same perturbation magnitude.

This increased sensitivity is attributed primarily to depth-dependent magnification effects and anatomical overlap along the lateral projection direction. As a result, small intrinsic calibration perturbations produced amplified reprojection displacement in LAT imaging geometry.

Figure~\ref{fig:landmark_shift} further demonstrates that landmark reprojection displacement is substantially larger in LAT views, particularly near posterior anatomical structures including the pedicles and spinous process.

Quantitative analysis across increasing perturbation magnitudes demonstrated consistently larger contour mismatch and reduced Dice similarity in LAT projections, confirming that LAT imaging geometry is more sensitive to intrinsic calibration variation than AP imaging geometry.

\subsection{2D--3D Registration Error Analysis}

\begin{figure*}[ht]
    \centering

    \begin{subfigure}{0.48\textwidth}
        \centering
        \includegraphics[width=\linewidth]{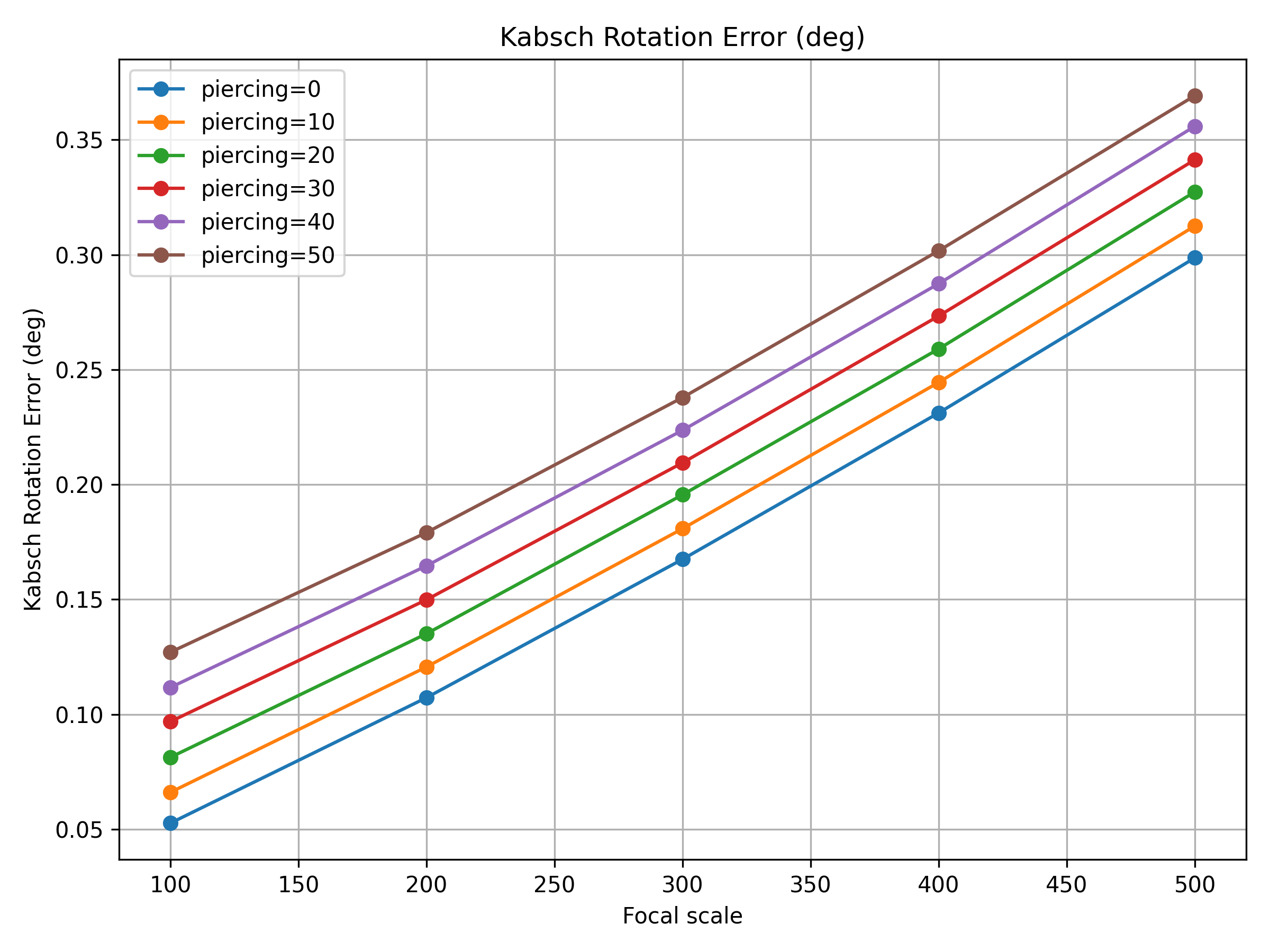}
        \caption{Kabsch rotation error.}
        \label{fig:kabsch_rotation}
    \end{subfigure}
    \hfill
    \begin{subfigure}{0.48\textwidth}
        \centering
        \includegraphics[width=\linewidth]{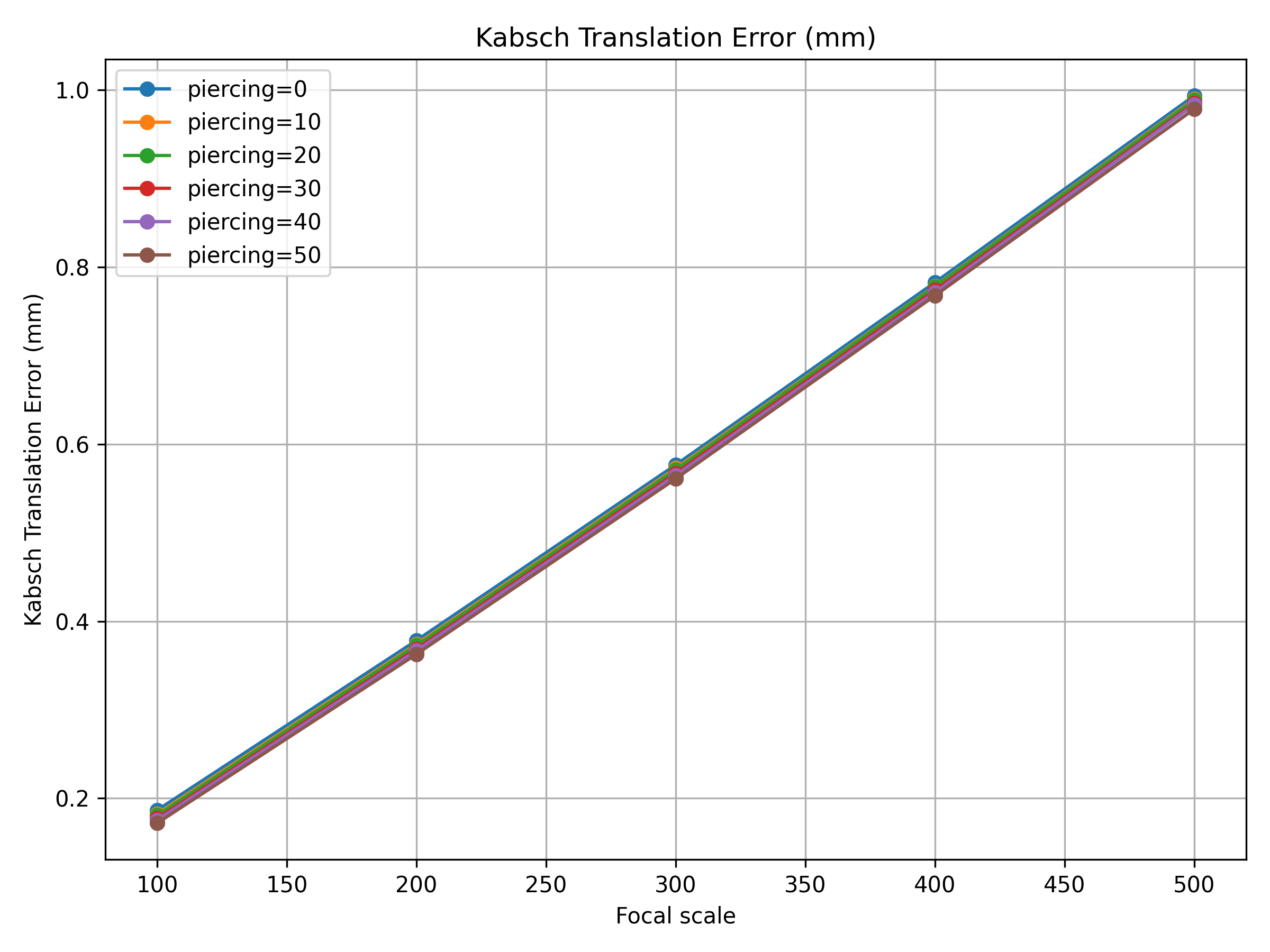}
        \caption{Kabsch translation error.}
        \label{fig:kabsch_translation}
    \end{subfigure}

    \caption{Kabsch-based registration error under varying focal scale perturbation and piercing conditions.}
    \label{fig:kabsch_combined}
\end{figure*}

Figure~\ref{fig:kabsch_rotation} and Figure~\ref{fig:kabsch_translation} summarize registration performance obtained using Kabsch-based rigid alignment under varying focal scale perturbations and piercing conditions. Rotation error increased progressively with increasing focal scale perturbation, demonstrating increasing sensitivity of rigid alignment to intrinsic calibration variation. For piercing values ranging from 0 to 50, Kabsch rotation error increased from approximately 0.05$^\circ$ to 0.37$^\circ$ across the evaluated focal scale range. In contrast, translational error remained comparatively stable, increasing from approximately 0.18~mm to 1.0~mm with minimal separation between piercing conditions.

\begin{figure*}[ht]
    \centering

    \begin{subfigure}{0.48\textwidth}
        \centering
        \includegraphics[width=\linewidth]{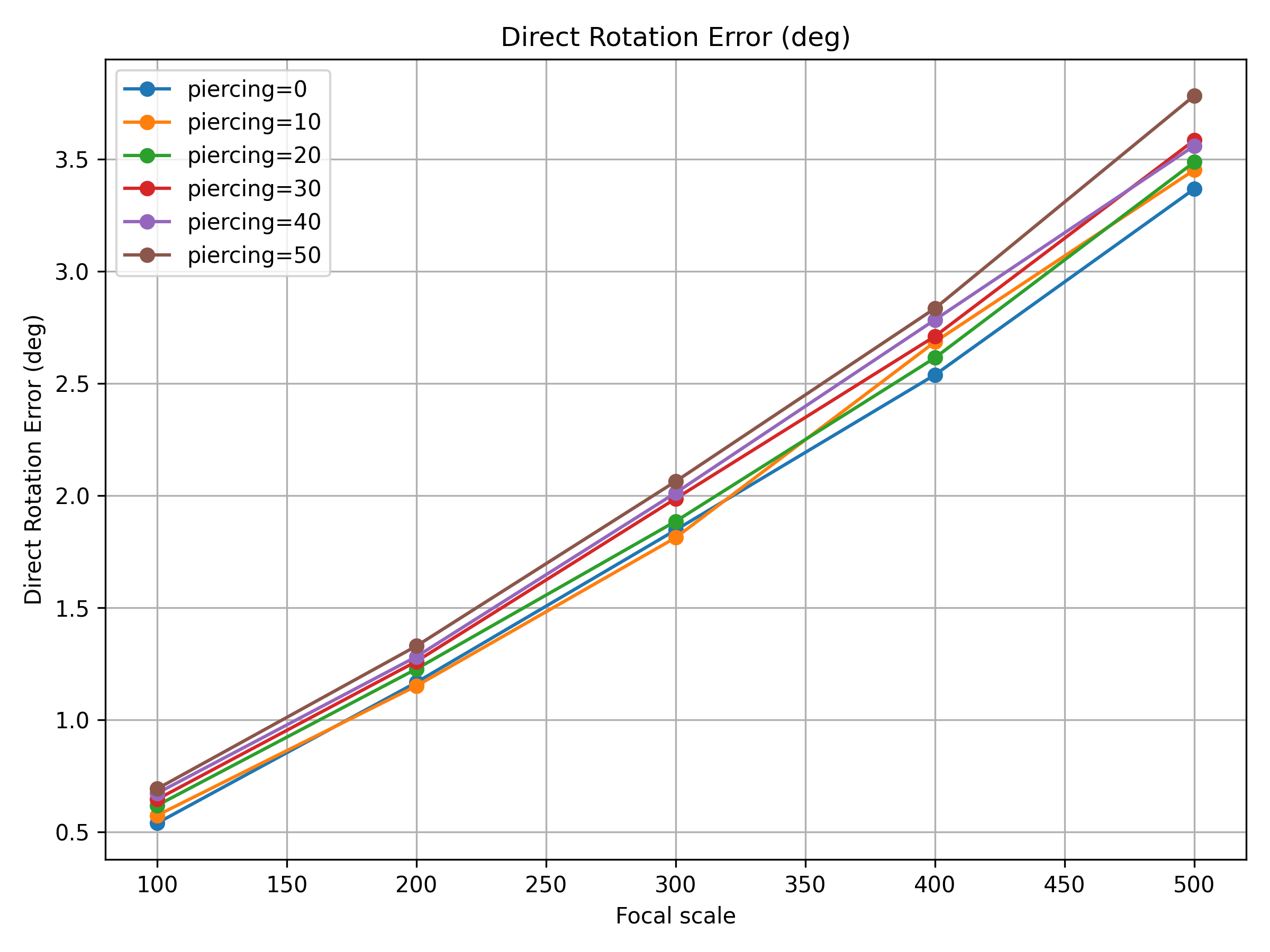}
        \caption{Direct registration rotation error.}
        \label{fig:direct_rotation}
    \end{subfigure}
    \hfill
    \begin{subfigure}{0.48\textwidth}
        \centering
        \includegraphics[width=\linewidth]{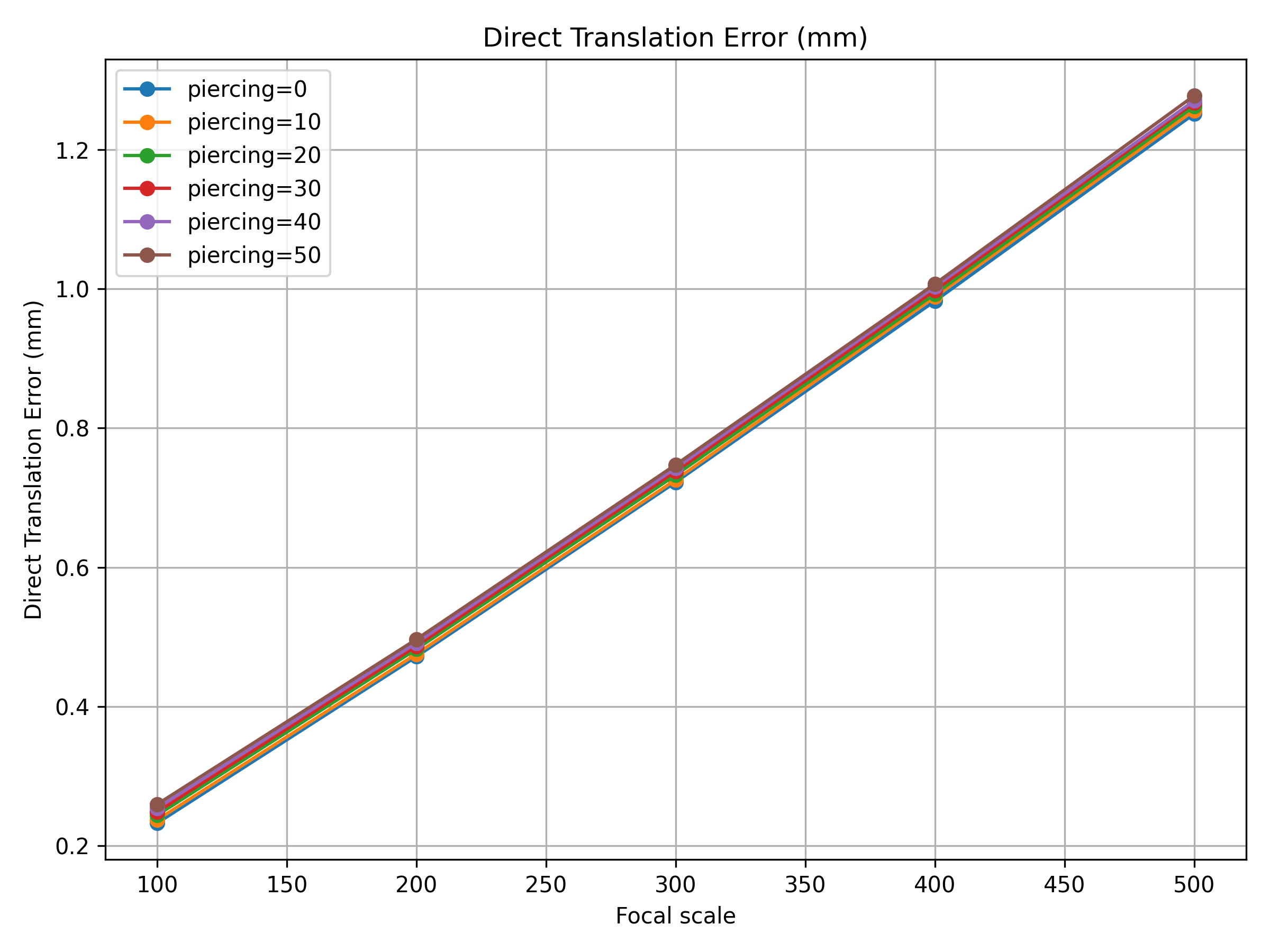}
        \caption{Direct registration translation error.}
        \label{fig:direct_translation}
    \end{subfigure}

    \caption{Direct registration error under varying focal scale perturbation and piercing conditions.}
    \label{fig:direct_combined}
\end{figure*}

Direct registration results are shown in Figure~\ref{fig:direct_rotation} and Figure~\ref{fig:direct_translation}. Direct rotational error exhibited substantially larger magnitude than Kabsch-based rotational error, increasing from approximately 0.5$^\circ$ to 3.8$^\circ$ as focal scale increased. Higher piercing conditions consistently produced larger rotational deviation, indicating amplified sensitivity to projection inconsistency under more challenging geometric configurations. Direct translational error also increased monotonically with focal scale perturbation, ranging from approximately 0.24~mm to 1.28~mm.

Overall, the experiments demonstrate that intrinsic calibration perturbation produces measurable degradation in downstream 2D--3D registration accuracy, particularly in rotational alignment metrics. The results further suggest that projection-domain inconsistency induced by calibration variation may propagate into rigid registration estimation even when anatomical geometry remains unchanged.

\section{Discussion}

This study investigated how intrinsic calibration perturbations influence vertebral projection appearance in synthetic fluoroscopic imaging. Using CT-derived vertebral models and controlled DRR generation, we evaluated how changes in intrinsic calibration parameters affect projection contours, silhouette overlap, and anatomical landmark localization under fixed AP and LAT imaging geometries.

The experimental results demonstrated measurable projection differences under intrinsic calibration perturbation, particularly in lateral imaging views. Although the underlying vertebral anatomy and imaging pose remained fixed, perturbations in focal length and related intrinsic parameters produced observable changes in contour shape and landmark projection location. These findings indicate that projection appearance may be sensitive to relatively small variations in imaging calibration.

Conventional fluoroscopic calibration methods are commonly evaluated using reconstruction-domain metrics such as reprojection error or reconstruction accuracy \cite{navab1998camera,siewerdsen2008cone}. While these metrics provide important measures of geometric consistency, they do not directly characterize projection-domain variation in synthesized fluoroscopic appearance. The present study therefore focused specifically on projection-domain analysis using DRR-based vertebral projections and contour similarity measurements.

The experiments additionally demonstrated clear view-dependent sensitivity differences between AP and LAT imaging
geometries. Across multiple perturbation magnitudes, LAT projections consistently exhibited larger contour deformation,
reduced silhouette overlap, and increased landmark displacement compared with AP projections. This behavior should
be interpreted in the context of the non-isocentric C-arm geometry used in this study. As described in
Section~\ref{sec:experimental_setup}, the vertebral model was assumed to remain fixed between AP and LAT views,
and no inter-view anatomical motion, repositioning, or deformation was introduced. Therefore, the observed AP/LAT
differences were not caused by vertebral motion, but instead reflected the combined effects of view-dependent imaging
geometry and intrinsic calibration perturbation.

In the non-isocentric setup, the source--object--detector relationship differed between AP and LAT configurations.
In particular, the vertebra was positioned closer to the detector in the LAT view and consequently appeared larger in
the projected image. Because the anatomy occupied a larger detector region in LAT, the same intrinsic calibration
perturbation produced larger apparent landmark displacement and contour deformation than in AP. In addition,
overlapping vertebral anatomy and depth compression in the lateral projection direction may further amplify local
projection-domain differences.

The landmark-based 2D--3D registration experiments further showed that projection-domain inconsistencies caused by intrinsic calibration perturbations can propagate into downstream pose estimation. When registration was performed using perturbed intrinsic calibration parameters, the recovered vertebral pose exhibited increased residual reprojection error and measurable degradation in registration accuracy compared with the ground-truth calibration condition. This effect was particularly evident in rotational alignment estimates, suggesting that calibration-induced projection changes may alter the optimization landscape and bias the estimated rigid transformation. These findings support the importance of considering intrinsic calibration uncertainty not only as an imaging geometry issue, but also as a potential source of error in DRR-based 2D--3D registration workflows.

An additional contribution of this work is the development of a controlled perturbation framework for projection-domain analysis. By introducing analytically defined intrinsic calibration perturbations under fixed imaging poses, the framework isolates projection changes associated specifically with calibration variation while minimizing confounding effects from motion or anatomical deformation. This enables systematic quantitative comparison of projection sensitivity using landmark displacement, contour distance, Dice similarity coefficient, and IoU measurements.

\subsection{Clinical and Technical Implications}

The presented findings suggest that projection-domain analysis may provide complementary information for evaluating fluoroscopic imaging geometry.

First, the observed projection differences indicate that intrinsic calibration variation can influence vertebral projection appearance even in controlled synthetic imaging conditions. Since many fluoroscopy-guided workflows rely on projection visualization and DRR generation, projection consistency may represent a useful additional consideration during calibration analysis.

Second, the increased sensitivity observed in LAT imaging suggests that projection behavior depends not only on
imaging orientation, but also on the source--object--detector geometry. In the present non-isocentric setup, the
vertebra was closer to the detector in the LAT configuration and therefore appeared larger in the image. This larger
projected scale made LAT projections more sensitive to intrinsic calibration perturbation. In spinal fluoroscopy,
lateral projections also contain compressed anatomical structures and overlapping vertebral features, which may
further increase sensitivity to geometric perturbation.

Third, the proposed framework provides a controlled environment for studying projection-domain variation in vertebral fluoroscopic imaging. The framework may be useful for future studies involving calibration analysis, DRR generation evaluation, or synthetic fluoroscopic simulation.

Potential applications of the proposed analysis framework include:

\begin{itemize}
    \item fluoroscopic projection consistency evaluation,
    \item DRR generation analysis,
    \item vertebral landmark projection analysis,
    \item synthetic fluoroscopic simulation studies,
    \item calibration sensitivity visualization.
\end{itemize}

\subsection{Limitations}

Several limitations of this study should be acknowledged.

First, the experiments were performed using synthetic fluoroscopic projections generated from CT-derived vertebral mesh models. Although this enables precise control of imaging geometry and perturbation magnitude, real fluoroscopic systems contain additional physical effects including scatter, detector noise, beam hardening, distortion, and nonuniform detector response that were not modeled in the current framework.

Second, the study focused on rigid vertebral anatomy and did not incorporate anatomical deformation, soft tissue interaction, or patient motion. These factors may further influence projection appearance in clinical imaging environments.

Third, the perturbation models were analytically defined and primarily limited to intrinsic calibration variation. Real imaging systems may involve more complex combinations of intrinsic, extrinsic, temporal, and system-dependent geometric variation.

Finally, although this study included a landmark-based 2D--3D registration experiment, the registration analysis was limited to a controlled synthetic setting with predefined anatomical landmarks. The current framework did not evaluate intensity-based registration, full clinical optimization pipelines, target registration error, or robustness under realistic fluoroscopic image degradation. Additional studies are therefore required to determine how the observed calibration-induced projection variations influence clinical 2D--3D registration workflows using real fluoroscopic images.

Future work will extend the proposed framework through:

\begin{itemize}
    \item validation using real fluoroscopic datasets,
    \item incorporation of realistic detector physics and noise models,
    \item deformable anatomical and motion-aware simulations,
    \item multi-view and multi-vertebral projection analysis,
    \item evaluation of additional calibration perturbation models.
\end{itemize}

\section{Conclusion}

This paper presented a synthetic framework for analyzing the sensitivity of vertebral fluoroscopic projections and landmark-based 2D--3D registration to intrinsic calibration perturbation. Using CT-derived vertebral models and digitally reconstructed radiographs generated under controlled imaging geometry, we quantitatively evaluated how intrinsic parameter variation influences projection-domain anatomical appearance in AP and LAT views.

Experimental results demonstrated that relatively small intrinsic calibration perturbations can produce measurable changes in vertebral contour geometry, landmark location, silhouette overlap, and DRR appearance despite fixed anatomy and imaging pose. Lateral projections consistently exhibited greater sensitivity to perturbation than AP projections, likely because of increased anatomical overlap and compressed depth geometry.

The landmark-based 2D--3D registration experiment further showed that calibration-induced projection inconsistencies can propagate into pose estimation error. Registration performed with perturbed intrinsic calibration produced increased residual reprojection error and degraded pose recovery compared with the ground-truth calibration condition, with rotational alignment being particularly sensitive. These results indicate that intrinsic calibration uncertainty can affect not only projection appearance but also downstream 2D--3D registration accuracy.

Overall, these findings suggest that projection-domain analysis provides complementary information beyond conventional reconstruction-based calibration evaluation and may improve understanding of geometric sensitivity in fluoroscopy-guided vertebral imaging systems. Future work will extend this framework to larger datasets, realistic fluoroscopic physics, intensity-based registration pipelines, and validation using real fluoroscopic images.

\bibliography{bio} % bibliography data in report.bib
\bibliographystyle{unsrt}

\end{document}